\documentclass[journal,twoside]{IEEEtran}
\usepackage{amsfonts}
\usepackage{amssymb}
\usepackage{mathrsfs}
\usepackage{verbatim}
\usepackage{subfigure}
\usepackage[centertags]{amsmath}
\usepackage{graphicx}

\newtheorem{definition}{Definition}
\newtheorem{lemma}{{ Lemma}}
\newtheorem{exmp}{Example}[section]
\newtheorem{theorem}{\bf Theorem}
\def\f{{\mathbb{F}}}
\usepackage{epsfig}
\ifCLASSINFOpdf
\else
\fi
\usepackage{array}

\begin{document}
%
\title{Square Complex Orthogonal Designs with Low PAPR and Signaling Complexity}
\author{Smarajit Das,~\IEEEmembership{Student Member,~IEEE} and
        B. Sundar Rajan,~\IEEEmembership{Senior Member,~IEEE}%

\thanks{This work was supported through grants to B.S.~Rajan; partly by the IISc-DRDO program on Advanced Research in Mathematical Engineering, and partly by the Council of Scientific \& Industrial Research (CSIR, India) Research Grant (22(0365)/04/EMR-II). The material in this paper was presented in parts at the IEEE International Symposium on Information theory held at Nice, France during June 24-29, 2007. Smarajit Das and B. Sundar Rajan are with the Department of Electrical Communication Engineering, Indian Institute of Science, Bangalore-560012, India. Email:\{smarajit,bsrajan\}@ece.iisc.ernet.in.}
\thanks{Manuscript received July 15, 2007; revised November 22, 2007.}}

%
%

\markboth{IEEE Transactions on Wireless Communications ,~Vol.~xx, No.~xx, xxxx}{Das \MakeLowercase{and} Rajan: Square Complex Orthogonal Designs with Low PAPR and Signaling Complexity}
%



\maketitle

\begin{abstract}
Space-Time Block Codes from square complex orthogonal designs (SCOD) have been extensively studied and most of the existing SCODs contain large number of zero. The zeros in the designs result in high peak-to-average power ratio (PAPR) and also impose a severe constraint on hardware implementation of the code when turning off some of the transmitting antennas whenever a zero is transmitted. Recently, rate $\frac{1}{2}$ SCODs with no zero entry have been reported for 8 transmit antennas. In this paper, SCODs with no zero entry for $2^a$ transmit antennas whenever $a+1$ is a power of $2$, are constructed which includes the $8$ transmit antennas case as a special case. More generally, for arbitrary values of $a$, explicit construction of $2^a\times 2^a$ rate $\frac{a+1}{2^a}$ SCODs with the ratio of number of zero entries to the total number of entries equal to $1-\frac{a+1}{2^a}2^{\lfloor log_2(\frac{2^a}{a+1}) \rfloor}$ is reported, whereas for standard known constructions, the ratio is $1-\frac{a+1}{2^a}$. The codes presented do not result in increased signaling complexity. Simulation results show that the codes constructed in this paper outperform the codes using the standard construction under peak power constraint while performing the same under average power constraint.

\end{abstract}


%

\section{Introduction}
\IEEEPARstart{S}{pace-Time} Block Codes (STBCs) from Complex Orthogonal Designs (CODs) have been extensively studied in \cite{TJC,TiH,Lia}. 

Let $x_1,x_2,\cdots,x_t$ be commuting, real indeterminates. A real orthogonal design $\mathbf{X}$ of order $n$ and type $(a_1,a_2,\cdots,a_t)$, denoted as $OD(n;a_1,a_2,\cdots,a_t)$ where the coefficients $a_i$ are positive integers, is a matrix of order $n$ with entries chosen from $0,\pm x_1,\pm x_2,\cdots,\pm x_t$, such that
$\mathbf{X}^{\mathcal{T}}\mathbf{X}=(a_1x_1^2 + a_2x_2^2+\cdots+ a_tx_t^2)\mathbf{I}_n$
where $\mathbf{X}^{\mathcal{T}}$ denotes the transpose of the matrix $X$ and $\mathbf{I}_n$ is the $ n \times n $ identity matrix.

Amicable orthogonal designs (AODs) are defined using two real orthogonal designs of same order but not necessarily of same type.
Let $\mathbf{X}$ be an $OD(n;u_1,u_2,\cdots,u_s)$ on the real variables $x_1,x_2,\cdots,x_s$
and let $\mathbf{Y}$ be an $OD(n;v_1,v_2,\cdots,v_t)$ on the real variables $y_1,y_2,\cdots,y_t$. It is said that $\mathbf{X}$ and $\mathbf{Y}$ are $AOD(n;u_1,u_2,\cdots,u_s;v_1,v_2,\cdots,v_t)$ if $\mathbf{X}\mathbf{Y}^{\mathcal{T}}=\mathbf{Y}\mathbf{X}^{\mathcal{T}}$.

 Amicable orthogonal designs have been studied by several authors \cite{ZSXWWWT,GeS} to construct complex orthogonal designs. The book by Geramita and Seberry \cite{GeS} gives a nice introduction to this topic.  

In the following, we define square complex orthogonal design which we use frequently in the rest of the paper.
A {\textit{Square Complex Orthogonal Design}} (SCOD) $\mathbf{G} (x_1, x_2, . . . , x_k )$  (in short $\mathbf{G}$) of size $n$ is an  $ n \times n $ matrix such that:
\begin{itemize}
\item the entries of $\mathbf{G} (x_1 , x_2 , . . . , x_k ) $ are complex linear combinations of the variables $ x_1 , x_2 , . . . , x_k $ and their complex conjugates $ x_1^* , x_2^* , . . . , x_k^*,$
\item $\mathbf{G}^{\mathcal{H}} \mathbf{G}=({\vert x_1 \vert}^2 + ...+{\vert x_k \vert}^2)\mathbf{I}_n$  where $\mathcal{H}$ stands for the complex conjugate transpose and $\mathbf{I}_n$ is the $ n \times n $ identity matrix.
\end{itemize}
If the non-zero entries are the indeterminates $ \pm x_1,\cdots$,$\pm x_k$ or their conjugates $ \pm x_1^*,\pm x_2^*,...,\pm x_k^* $ only (not arbitrary complex linear combinations), then $\mathbf{G}$ is said to be a {\textit {restricted complex orthogonal design}} (RCOD). The rate of $\mathbf{G}$ is $\frac{k}{n}$ complex symbols per channel use.

It is known that the maximum rate $\mathcal{R}$ of an  $n\times n$ RCOD is $\frac{a+1}{n}$ where $n=2^a(2b+1), a\mbox{ and } b$ are positive integers \cite{TiH}.
Note that the maximal rate does not depend on $b.$ Several authors have constructed RCODs for $2^a$ antennas achieving maximal rate \cite{TiH,ALP,Joz,Wol}. In \cite{TiH}, the following induction method is used to construct SCODs for $2^a$ antennas, $a=2,3,\cdots$, starting from 
\begin{equation}
\label{itcod}
\mathbf{G}_1= \left[ 
\begin{array}{ c @{\hspace{.2pt}} c @{\hspace{.2pt}}}
x_1   &-x_2^*      \\
x_2   & x_1^*
\end{array}\right],~
\mathbf{G}_a=  \left[ 
\begin{array}{ c @{\hspace{.4pt}} c @{\hspace{.2pt}}}
\mathbf{G}_{a-1}   & -x_{a+1}^*\mathbf{I}_{2^{a-1}}      \\
x_{a+1}\mathbf{I}_{2^{a-1}}   & \mathbf{G}_{a-1}^{\mathcal{H}}
\end{array}\right]
\end{equation} 
\noindent
where $\mathbf{G}_a$ is a $2^a\times 2^a$ complex matrix. Note that $\mathbf{G}_a$ is a RCOD in $a+1$ complex variables $x_1, x_2,\cdots, x_{a+1}$. Moreover, each row and each column of the matrix $\mathbf{G}_a$ contains only $a+1$ non-zero elements and all other entries in the same row or column are filled with zeros. The fraction of zeros, defined as the ratio of the number of zeros to the total number of entries in a design, for $\mathbf{G}_a$, is
\begin{equation}
\label{largefrac}
\frac{2^a-a-1}{2^a}=1-\frac{a+1}{2^a}=1-\mathcal{R}.
\end{equation}
For the constructions in  \cite{TiH,ALP,Joz,Wol} also, the fraction of zeros is the same as given by \eqref{largefrac}.  Reducing number of zeros in a SCOD for more than $2$ transmit antennas (for two antennas, the Alamouti code does not have any zeros), is important for many reasons, namely improvement in Peak-to-Average Power Ratio (PAPR) and also the ease of practical implementation of these codes in wireless communication system \cite{YGT1}.

\begin{figure*}
{\tiny
\begin{equation*}
\label{TWMS8code}
\mathbf{G}_{TWMS}=
\frac{1}{\sqrt{2}}\left[
\begin{array}{r@{\hspace{0.9pt}}r@{\hspace{0.9pt}}r@{\hspace{0.9pt}}r@{\hspace{0.9pt}}r@{\hspace{0.9pt}}r@{\hspace{0.9pt}}r@{\hspace{0.9pt}}r@{\hspace{0.9pt}}}
x_{1} &x_{1} &x_{2} &x_{2} &x_{3} &x_{4} &x_{3} &x_{4}  \\
x_{1} &-x_{1} &x_{2} &-x_{2} &x_{4}^* &-x_{3}^* &x_{4}^* &-x_{3}^*  \\
x_{2}^* &x_{2}^* &-x_{1}^* &-x_{1}^* &x_{3} &x_{4} &-x_{3} &-x_{4}  \\
x_{2}^* &-x_{2}^* &-x_{1}^* &x_{1}^* &x_{4}^* &-x_{3}^* &-x_{4}^* &x_{3}^*  \\
x_{4I}+jx_{3Q}  &x_{3I}+jx_{4Q} &x_{4I}+jx_{3Q} &x_{3I}+jx_{4Q} &x_{2I}+jx_{1Q} &x_{2I}+jx_{1Q} &x_{1I}+jx_{2Q} &x_{1I}+jx_{2Q}  \\
x_{3I}+jx_{4Q}  &x_{4I}+jx_{3Q} &x_{3I}+jx_{4Q} &x_{4I}+jx_{3Q} &x_{2I}+jx_{1Q} &x_{2I}+jx_{1Q} &x_{1I}+jx_{2Q} &x_{1I}+jx_{2Q}  \\
x_{4I}+jx_{3Q}  &x_{3I}+jx_{4Q} &x_{4I}+jx_{3Q} &x_{3I}+jx_{4Q} &x_{1I}+jx_{2Q} &x_{1I}+jx_{2Q} &x_{2I}+jx_{1Q} &x_{2I}+jx_{1Q}  \\
x_{3I}+jx_{4Q}  &x_{4I}+jx_{3Q} &x_{3I}+jx_{4Q} &x_{4I}+jx_{3Q} &x_{1I}+jx_{2Q} &x_{1I}+jx_{2Q} &x_{2I}+jx_{1Q} &x_{2I}+jx_{1Q}
\end{array}
\right]
\end{equation*}
}
\hrule
\end{figure*}

For illustration, consider the SCOD $\mathbf{G}_2$ of size $4$ shown below - it is  a RCOD,  whereas the code $\mathbf{G}_{TJC}$ also shown below, given in ~\cite{TJC,GeS}, obtained from Amicable Orthogonal Designs, is not a RCOD and there are no zeros in this matrix. 

{\small
\begin{eqnarray}
\label{itcod4}
\begin{array}{c}
\mathbf{G}_2=\left[
\begin{array}{rrrr}
x_1   & -x_2^*   &-x_3^*   &0    \\
x_2   & x_1^* &0     &-x_3^*   \\
x_3 & 0     &x_1^* & x_2^* \\
0    &x_3   &-x_2  &x_1 
\end{array}
\right],\\

\mathbf{G}_{TJC}=\left[
\begin{array}{ c @{\hspace{.4pt}} c @{\hspace{.2pt}}c @{\hspace{.2pt}}c @{\hspace{.2pt}}}
x_1   & x_2   &\frac{x_3}{\sqrt{2}} & \frac{x_3}{\sqrt{2}}     \\
-x_2^* & x_1^* &\frac{x_3}{\sqrt{2}} & \frac{-x_3}{\sqrt{2}}   \\
\frac{x_3^*}{\sqrt{2}} & \frac{x_3^*}{\sqrt{2}}  &\frac{(-x_1-x_1^*+x_2-x_2^*)}{2}& \frac{(x_1-x_1^*-x_2-x_2^*)}{2} \\
\frac{x_3^*}{\sqrt{2}} & \frac{-x_3^*}{\sqrt{2}}  &\frac{(x_1-x_1^*+x_2+x_2^*)}{2} &-\frac{(x_1+x_1^*+x_2-x_2^*)}{2}
\end{array}
\right]
\end{array}
\end{eqnarray}
}
Notice that some of the entries of $\mathbf{G}_{TJC}$ can be written as

\begin{equation}
\label{ciod}
\begin{array}{rrr}
\frac{(-x_1-x_1^*+x_2-x_2^*)}{2}=& -(x_{1I}-jx_{2Q})=& -{\hat x}_1^*,\\ 
\frac{(x_1-x_1^*-x_2-x_2^*)}{2}=& -(x_{2I}-jx_{1Q})=& -{\hat x}_2^*, \\
\frac{(x_1-x_1^*+x_2+x_2^*)}{2}=&x_{2I}+jx_{1Q}=&{\hat x}_2, \\ 
-\frac{(x_1+x_1^*+x_2-x_2^*)}{2}=&-(x_{1I}+jx_{2Q})=& -{\hat x}_1,
\end{array}
\end{equation}
where ${\hat x}_1=x_{1I}+jx_{2Q}$ and ${\hat x}_2=x_{2I}+jx_{1Q}$ are the coordinate interleaved variables corresponding to the variables $x_1$ and $x_2,$ where $x_{iI}$ and $x_{iQ}$ are the in-phase and the quadrature-phase of the  variable $x_i$. Single-Symbol ML Decodable Designs based on coordinate interleaved variables have been studied in \cite{KhR}. For our purposes, it is important  to note that whenever coordinate interleaving appears, it is nothing but a specific complex linear combination of two variables, which will have impact in terms of the signaling complexity explained subsequently.\\ 
The following code $\mathbf{G}_3$ for $8$ transmit antennas,

\begin{eqnarray}
\label{itcod8}
\begin{array}{c}
\mathbf{G}_3=\left[\begin{array}{r@{\hspace{0.9pt}}r@{\hspace{0.9pt}}r@{\hspace{0.9pt}}r@{\hspace{0.9pt}}r@{\hspace{0.9pt}}r@{\hspace{0.9pt}}r@{\hspace{0.9pt}}r@{\hspace{0.9pt}}}
    x_1   &-x_2^*  &-x_3^*& 0    &-x_4^* & 0     & 0    & 0 \\
    x_2   & x_1^*  & 0    &-x_3^*& 0     &-x_4^* & 0    & 0 \\
    x_3   & 0      & x_1^*& x_2^*& 0     & 0     &-x_4^*& 0 \\
     0    & x_3    &-x_2  & x_1  & 0     & 0     & 0    &-x_4^* \\
    x_4   & 0      & 0    & 0    & x_1^* & x_2^* & x_3^*& 0 \\
     0    & x_4    & 0    & 0    &-x_2   & x_1   & 0    & x_3^*\\
     0    & 0      & x_4  & 0    &-x_3   & 0     & x_1  &-x_2^*\\
     0    & 0      & 0    & x_4  & 0     &-x_3   & x_2  & x_1^*\\
\end{array}\right],\\
\mathbf{G}_Y=\left[\begin{array}
{r@{\hspace{0.4pt}}r@{\hspace{0.4pt}}r@{\hspace{0.4pt}}r@{\hspace{0.4pt}}r@{\hspace{0.4pt}}r@{\hspace{0.4pt}}r@{\hspace{0.4pt}}r@{\hspace{0.4pt}}}
x_{1}^*  & x_{1}^*& x_{2}   &-x_{2}   &x_{3}   &-x_{3}  & x_{4}  & -x_{4}   \\
jx_{1}   & -jx_{1}  & jx_{2}^* &jx_{2}^* &jx_{3}^* & jx_{3}^*& jx_{4}^*& jx_{4}^* \\
-x_{2}   & x_{2}  & x_{1}^* &x_{1}^* &x_{4}^* & -x_{4}^*& -x_{3}^*& x_{3}^* \\
-jx_{2}^*& -jx_{2}^*& jx_{1}   &-jx_{1}   &jx_{4}   & jx_{4}  & -jx_{3}  & -jx_{3}   \\
-x_{3}   & x_{3}  & -x_{4}^* &x_{4}^* &x_{1}^* & x_{1}^*& x_{2}^*& -x_{2}^*   \\
-jx_{3}^*& -jx_{3}^*& -jx_{4}   &-jx_{4}   &jx_{1}   & -jx_{1}  & jx_{2}  & jx_{2}   \\
-x_{4}   & x_{4}  & x_{3}^* &-x_{3}^* &-x_{2}^* & x_{2}^*& x_{1}^*& x_{1}^*   \\
-jx_{4}^*& -jx_{4}^* & jx_{3}  &jx_{3}   &-jx_{2}   & -jx_{2}  & jx_{1}  & -jx_{1}
\end{array}\right]
\end{array}
\end{eqnarray}
\noindent
contains 50 per cent of entries zeros. But, Yuen et al, in \cite{YGT}, have constructed a new rate-$1/2,$ SCOD $\frac{\mathbf{G}_Y}{\sqrt{2}}$ of size $8$ with no zeros in the design matrix using Amicable Complex Orthogonal Design (ACOD) \cite{GeS} where $\mathbf{G}_Y$ is given in \eqref{itcod8}.
 
Observe that for a fixed average power per codeword, due to the presence of zeros in $\mathbf{G}_3$, the peak power transmission in an antenna using $\mathbf{G}_3$ will be higher than that of an antenna using $\mathbf{G}_Y.$  Hence, it is clear that the PAPR for the code $\mathbf{G}_Y$ is lower than that of the code $\mathbf{G}_3.$ Hence,  lower the fraction of zeros in a code, lower will be the PAPR of the code. In \cite{TWMS,STWWWXZ,ZSXWWWT}, another rate-$1/2,$ $8$ antenna code with no zero entry, denoted by $\mathbf{G}_{TWMS}$ shown at the top of this page, has been reported.

Observe that $\mathbf{G}_{TWMS}$ has entries that are coordinated interleaved variables and hence has larger signaling complexity.

\subsection*{Signaling complexity:}
Notice that some of the entries, for instance $x_{1I}+jx_{2Q}$ and $x_{2I}+jx_{1Q}$, in $\mathbf{G}_{TJC}$ and $\mathbf{G}_{TWMS}$ are co-ordinate interleaved versions of the variables $x_1$ and $x_2$. Suppose the variables $x_1$ and $x_2$ take values from a regular 
(rectangular) 16-QAM rotated by an angle $\theta$. Though rotation does not affect the full-diversity of the code, the coding gain depends on $\theta$ and hence non-zero value of $\theta$ may be desired. Now the antenna transmitting $x_1$ chooses one of the $16$ complex numbers for transmission whereas the antenna transmitting $x_{1I}+jx_{2Q}$ will be choosing one of $16 \times 16$ complex numbers since the components $x_{1I}$ and $x_{2Q}$ take independently $16$ values each. This will increase the number of quantization levels needed in a digital implementation for signals transmitted in this antenna. We will henceforth refer to the number of quantization levels needed in such a digital implementation as ``signaling complexity''. Notice that designs which have entries that are linear combinations of several variables increase the signaling complexity of the design.
Accordingly, the signaling complexity of $G_2$ given in \eqref{itcod4} is less than that of the code on the right hand side of \eqref{itcod4}. Similarly, the signaling complexity of $\mathbf{G}_{TWMS}$ is larger than that of $\mathbf{G}_Y$. 

Notice that by multiplying the matrix $\mathbf{G}_3$ with a unitary matrix, the resulting matrix will continue to be a SCOD with different number of zeros and it is not difficult to find unitary matrices that will result in a design with no zero entries. However, such a design is likely to have a large signaling complexity which needs to be avoided. Obtaining  a unitary matrix which reduces the number of zero entries while not increasing the signaling complexity is a nontrivial task which is the subject matter of this paper.

In this paper, we provide a general procedure to construct SCODs with fewer number of zeros compared to known constructions for any power of two number of antennas (greater than 4), without increasing the signaling complexity.   Our contributions are summarized as follows:
\begin{itemize}
\item Maximal-rate SCODs with no zero entry and minimum signaling complexity for $2^a$ transmit antennas whenever $a+1$ is a power of $2$, are constructed which includes the $8$ transmit antennas case as a special case. This matches with the construction given in \cite{YGT} for $8$ transmit antennas and beats the codes in \cite{TWMS,STWWWXZ,ZSXWWWT} for $8$ transmit antennas in terms of signaling complexity.
\item More generally, for arbitrary values of $a$, explicit construction of $2^a\times 2^a$,  rate $\frac{a+1}{2^a}$ SCODs with the ratio of number of zero entries to the total number of entries equal to $1-\frac{a+1}{2^a}2^{\lfloor log_2(\frac{2^a}{a+1}) \rfloor}$ is reported. Note that when $a+1$ is a power of two, our codes have no zeros. When  $a+1$ is not a power of two, it is conjectured that SCODs with smaller fraction of zero entries with rate $\frac{a+1}{2^a}$ and same signaling complexity do not exist. Our construction gives fewer number of zero entries compared to the well known constructions in \cite{TiH,ALP,Joz,Wol}.
\item Our construction is based on simple premultiplication of the code in \eqref{itcod} by a scaled unitary matrix consisting of only $+1,-1$ or $0$, whereas the constructions in \cite{YGT,TWMS,STWWWXZ,ZSXWWWT} depend on the existence and availability of AODs \cite{GeS}.
\item A general procedure to obtain the scaled unitary matrix that leads to a SCOD with small number of zero entries is given.
\item It is shown that the new codes presented in this paper admit a recursive relation similar to that admitted by $\mathbf{G}_a.$
\end{itemize}

The remaining content of the paper is organized as follows: In Section \ref{sec2}, we prove the main result of the paper given by Theorem \ref{fracz}. In Section \ref{sec3}, we give a procedure to compute the premultiplying matrix using which we can get the SCODs of this paper straightaway from the well-known construction given by \eqref{itcod}. The PAPR of the new codes constructed is discussed in Section \ref{sec4}. Simulation results are given in Section \ref{sec5}. A brief summary and a conjecture constitute Section \ref{sec6}.

\section{Construction of SCODs with Low PAPR}
\label{sec2}
 SCODs given in~\cite{TiH} contain a large number of zeros  and the fraction of zeros in the code increases as the number of transmit antenna increases. Note that these codes are RCODs and hence of least decoding complexity as well as least signaling complexity. It is possible to  obtain an orthogonal matrix with fewer zero, if we premultiply and/or post-multiply the given orthogonal design matrix by some unitary matrix, but the resulting orthogonal design need not be a RCOD. So care must be taken in how we choose these premultiplying or post-multiplying matrices such that the code obtained after applying these matrices, does not contain complex linear combination of the variables which will increase the signaling complexity.

There exists a unitary matrix which when pre-multiplies the code $\mathbf{G}_3$  obtain a code which contains no zero in the matrix and none of the entries of this new code is a complex linear combination of variables and thus the signaling complexity is not increased.
The unitary matrix corresponding to $\mathbf{G}_3$ is $\frac{1}{\sqrt{2}}\mathbf{Q}^{(3)}$ where $\mathbf{Q}^{(3)}$ is given by the matrix on the left hand side of \eqref{matrix8}. Here $-1$ is represented by simply the minus sign (throughout the paper) and the resulting no zero entry SCOD is $\mathbf{H}_3$ where the matrix $\sqrt{2}\mathbf{H}_3$ is shown on the right hand side of \eqref{matrix8}.

\begin{eqnarray}
\label{matrix8}
\left[\begin{array}
{r@{\hspace{0.2pt}}r@{\hspace{0.2pt}}r@{\hspace{0.2pt}}r@{\hspace{0.2pt}}r@{\hspace{0.2pt}}r@{\hspace{0.2pt}}r@{\hspace{0.2pt}}r@{\hspace{0.2pt}}}
  1&0&0&0&0&0&0&1 \\
  0&1&0&0&0&0&1&0  \\
  0&0&1&0&0&1&0&0  \\
  0&0&0&1&1&0&0&0  \\
  0&0&0&1&-&0&0&0 \\
  0&0&1&0&0&-&0&0 \\
  0&1&0&0&0&0&-&0 \\
  1&0&0&0&0&0&0&-
\end{array}\right],
\left[\begin{array}
{r@{\hspace{0.2pt}}r@{\hspace{0.2pt}}r@{\hspace{0.2pt}}r@{\hspace{0.2pt}}r@{\hspace{0.2pt}}r@{\hspace{0.2pt}}r@{\hspace{0.2pt}}r@{\hspace{0.2pt}}}
    x_1  &-x_2^*  &-x_3^*& x_4  &-x_4^*&-x_3   & x_2  & x_1^* \\
    x_2  & x_1^*  & x_4  &-x_3^*&-x_3  &-x_4^* & x_1  &-x_2^* \\
    x_3  & x_4    & x_1^*& x_2^*&-x_2  & x_1   &-x_4^*& x_3^* \\
    x_4  & x_3    &-x_2  & x_1  & x_1^*& x_2^* & x_3^*&-x_4^* \\
   -x_4  & x_3    &-x_2  & x_1  &-x_1^*&-x_2^* &-x_3^*&-x_4^* \\
    x_3  &-x_4    & x_1^*& x_2^*& x_2  &-x_1   &-x_4^*&-x_3^*\\
    x_2  & x_1^*  &-x_4  &-x_3^*& x_3  &-x_4^* &-x_1  & x_2^*\\
    x_1  &-x_2^*  &-x_3^*&-x_4  &-x_4^*&x_3    &-x_2  &-x_1^*
\end{array}\right]
\end{eqnarray}

Towards identifying such premultiplying matrices for the general case, we label the rows of  $\mathbf{G}_a$ as $R_0, R_1,\cdots,R_{2^a-1}$. The column index also varies from $0$ to $2^a-1$. Let $N_i^{(a)}$ be the set of column indices of the non-zero entries of the $i$-th row $R_i$ of the matrix $\mathbf{G}_a$.
The following lemma describes $N_i^{(a)}$ for all $i=0$ to $2^a-1$.

\begin{lemma}
Let $a$ be a positive integer and $\mathbf{G}_a$ be a COD of size $2^a\times 2^a$ in $(a+1)$ complex variables $x_1,\cdots, x_{a+1}$ as  given in \eqref{itcod}. Let $i$ be a positive integer between $0$ and $2^a-1$. Let the radix-2 representation of $i$ be $(i_{a-1}, i_{a-2},\cdots, i_0)$ where $i_{a-1}$ is the most significant bit.
Then
$$N_i^{(a)}=\{i\}\cup\{i+ (-1)^{i_j}2^j~\vert~ j=0,\cdots, a-1\} $$
or equivalently,
$N_i^{(a)}=\{i\}\cup\{i \oplus 2^j~\vert~ j=0 \mbox { to } a-1\} $ where $\oplus$ denotes the component-wise module 2 addition of the radix-2 representation vectors.
\end{lemma}
\begin{IEEEproof}
The proof is by induction on $a$. The case $a=1$, corresponds to the Alamouti code $\mathbf{G}_1$.
We note that $N_0^{(1)}=\{0,1\}$ and $N_1^{(1)}=\{0,1\}$ as given by the expression of $N_i^{(1)}$ for $i=0,1$.
 So for $a=1$, the lemma is true.
Let the lemma be true for all $a\leq n$.
Then, we have
\begin{equation}
\label{given}
N_i^{(n)}=\{i\}\cup\{i+ (-1)^{i_j}2^j\vert j=0,\cdots, n-1\} 
\end{equation}
for all $i=0,1,\cdots, 2^n-1$ and we need to prove that
\begin{equation}
\label{toprove}
N_i^{(n+1)}=\{i\}\cup \{i+ (-1)^{i_j}2^j\vert j=0,\cdots, n\}
\end{equation}
for all $i=0,1,\cdots, 2^{n+1}-1.$  For $a=n+1$, we have the radix-2 representation, $i=(i_n, i_{n-1},\cdots, i_0)$ and
\begin{eqnarray}
\label{bigmatrix}
\mathbf{G}_{n+1}=  \left[\begin{array}{rr}
   \mathbf{G}_n   & -x_{n+2}^*\mathbf{I}_{2^n}      \\
x_{n+2}\mathbf{I}_{2^{n}}   & \mathbf{G}_n^{\mathcal{H}}
   \end{array}\right].
\end{eqnarray}

We have the following two cases:

{\bf Case (i)  $ 0 \leq i \leq 2^n-1:$}
In this case $i_n=0$ and the term $i+(-1)^{i_n}2^{n}$  in \eqref{toprove} corresponds to the non-zero location in the $-x^*_{n+2} \mathbf{I}_{2^n}$ part of $\mathbf{G}_{n+1}$ and the nonzero locations in the $\mathbf{G}_n$ part is given by the remaining elements of \eqref{toprove} which is nothing but $N_i^{(n)}.$

{\bf Case (ii)  $ 2^n \leq i \leq 2^{n+1}-1:$}
In this case $i_n=1$ for all values of $i$ in the range under consideration. Then, the term corresponding to $j=n$ in \eqref{toprove} is $i-2^n$ which corresponds to the non-zero term in the $x_{n+2}\mathbf{I}_{2^n}$ part of the matrix \eqref{bigmatrix}. Also, every term of the form $i+(-1)^{i_j}2^j; j=0,1,\cdots, n-1$ in \eqref{toprove} will be same as a term in \eqref{given} with $2^n$ added to it. This takes into account all the non-zero entries in the $\mathbf{G}_n^{\mathcal{H}}$ part of \eqref{bigmatrix}.
\end{IEEEproof}

\begin{table*}
\caption{$M_a$  and $M_a^\prime$ for $a=3, \cdots 9$ } 
\begin{center}
\begin{tabular}{|c|c|c|c|c|c|c|c|} \hline \label{tab1}
$a$ & 3 & 4& 5& 6& 7& 8  & 9  \\ \hline
$M_a$ &$\{3\}$ &$\{3\}$&$\{3,5\}$&$\{3,5,6\}$ &$\{3,5,6,7\}$& $\{3,5,6,7\}$& $\{3,5,6,7,9\}$  \\ \hline
$M_a^\prime$ &$\{7\}$ & $\{7\}$&$\{7,25\}$&$\{7,25,42\}$ &$\{7,25,42,75\}$&$\{7,25,42,75\}$ &$\{7,25,42,75,385\}$ \\ \hline
$d$ &2 & 3&3&3 &3&4&4\\ \hline
\end{tabular}
\end{center}
\end{table*}

\begin{exmp}
In this example we compute the sets $N_i^{(a)}$ for $a=2$ and $3$. For $a=2$, the possible values of $i$ are $0,1,2$ and $3$, while for $a=3$, $i$ takes value between $0$ and $7$.
\begin{eqnarray*}
\begin{array}{c}
N_0^{(2)}=\{0\}\cup\{0\oplus2^0, 0\oplus 2^1\}=\{0,1,2\}, \\
N_1^{(2)}=\{1\}\cup\{1\oplus2^0, 1\oplus 2^1\}=\{1,0,3\}, \\
N_2^{(2)}=\{2\}\cup\{2\oplus2^0, 2\oplus 2^1\}=\{2,3,0\}, \\
N_3^{(2)}=\{3\}\cup\{3\oplus2^0, 3\oplus 2^1\}=\{3,2,1\}. \\
\end{array} \\ 
\begin{array}{c}
N_0^{(3)}=\{0\}\cup\{1,2,4\}=\{0,1,2,4\}, \\
N_1^{(3)}=\{1\}\cup\{0,3,5\}=\{1,0,3,5\}, \\
N_2^{(3)}=\{2\}\cup\{3,0,6\}=\{2,3,0,6\}, \\
N_3^{(3)}=\{3\}\cup\{2,1,7\}=\{3,2,1,7\}, \\
N_4^{(3)}=\{4\}\cup\{5,6,0\}=\{4,5,6,0\}, \\
N_5^{(3)}=\{5\}\cup\{4,7,1\}=\{5,4,7,1\}, \\
N_6^{(3)}=\{6\}\cup\{7,4,2\}=\{6,7,4,2\}, \\
N_7^{(3)}=\{7\}\cup\{6,5,3\}=\{7,6,5,3\}.
\end{array}
\end{eqnarray*}
Notice that $N_i^{(3)}\cap N_j^{(3)}=\phi$ if $i\oplus j=7,$ where $\phi$ represents the empty set.
\end{exmp}
\begin{definition}
Two rows $R_i,R_j$ of $\mathbf{G}_a$ are said to be {\it{non-intersecting}} if $N_i^{(a)}\cap N_j^{(a)}=\phi$.
\end{definition}

The following lemma is needed to prove Lemma \ref{part} which in turn is used in the proof of the main result given in Theorem \ref{fracz}.
\begin{lemma}
\label{dist3}
Let $V_a$ denote the set of all radix-2 representation vectors of the elements of the set $\{0,1,\cdots, 2^a-1 \}$ and $S$ be a subset of $V_a$. Then $N_i^{(a)}\cap N_j^{(a)}=\phi$ for all $i,j\in S, i\neq j$ if and only if the minimum Hamming distance (MHD) of $S$ is greater than or equal to $3$.
\end{lemma}
\begin{IEEEproof}
We first show that  $N_i^{(a)}\cap N_j^{(a)}=\phi$ for all $i,j\in S, i\neq j$ implies that the MHD of $S$ is greater than or equal to $3$. Equivalently,
if the MHD of $S$ is $1$ or $2$, then there exists $i, j\in S, i\neq j$, such that $N_i^{(a)}\cap N_j^{(a)}\neq\phi$. 
Assume that the MHD of $S$ is $1$ or $2$, then their exists $i,j\in S$, $i\neq j$, such that $dist (i,j)=1 \mbox{ or } 2$. So, either $i=j\oplus 2^k$ for some \\
$k\in\{0,1,\cdots, a-1\}$ if $dist(i,j)=1$, or $i=j\oplus 2^{k_1}\oplus 2^{k_2}$ for some $k_1,k_2\in \{0,1,\cdots, a-1\}, k_1\neq k_2$ if $dist(i,j)=2$. 
In the first case, $i\in N_i^{(a)}\cap N_j^{(a)}$ and in the second case, $(i\oplus 2^{k_1})\in N_i^{(a)}\cap N_j^{(a)}$ as $i\oplus 2^{k_1}=j \oplus 2^{k_2}$.\\
 For both cases, $N_i^{(a)}\cap N_j^{(a)}\neq\phi$. \\
%
%
Next we prove that if the MHD of $S$ is at least $3$, then $N_i^{(a)}\cap N_j^{(a)}=\phi$ for all $i,j\in S; i\neq j$, or equivalently, if for some $i, j\in S, i\neq j$, $N_i^{(a)}\cap N_j^{(a)}\neq\phi$, then MHD of $S$ is less than 3.\\
 Let $i,j\in S$ and $i\neq j$. 
We have\\
$N_i^{(a)}=\{i\oplus 2^k\vert k=0,\cdots, a-1\}\cup \{i\}$ and \\
$N_j^{(a)}=\{j\oplus 2^k\vert k=0,\cdots, a-1\}\cup \{j\}$.\\
As $N_i^{(a)}\cap N_j^{(a)}\neq \phi$, let $x\in N_i^{(a)}\cap N_j^{(a)}$.\\
We have $x=i$ or $x=i\oplus 2^{k_1}$ for some $0\leq k_1\leq a-1$, as $x\in N_i^{(a)}$.\\
Similarly,
 $x=j$ or $x=j\oplus 2^{k_2}$  for some $0\leq k_2\leq a-1$, as $x\in N_j^{(a)}$.\\
But if $x=i$, then $x\neq j$, as $i\neq j$.
So, we have following three cases:\\
(i) $x=i$ and $x=j\oplus 2^{k_2}$,\\
(ii) $x=i\oplus 2^{k_1}$ and $x=j$,\\
(iii) $x=i\oplus 2^{k_1}$ and $x=j \oplus 2^{k_2}$, $k_1\neq k_2$ (as $i\neq j$).\\
For the case (i) \& (ii), we have $i=j\oplus 2^{k_2}$ \& $i\oplus 2^{k_1}=j$ respectively and in both cases, $dist(i,j)=1$. For the case (iii), we have $i\oplus 2^{k_1}=j\oplus 2^{k_2}$, which means the $dist(i,j)=2$.
So MHD of $S$ is less than 3.
\end{IEEEproof}

For a given $a,$  let $d$ be the  positive integer such that $ 2^{d-1} \leq a < 2^d $ and $a= \sum_{j=0}^{d-1}a_j2^j,a_j\in \mathbb{F}_2$. Note that $a_{d-1}=1$. Define
\begin{equation}
\label{ma}
M_a=\{0<x\leq a~\vert ~ x\neq 2^k \text { for any $k=0,1,\cdots$ }\}
\end{equation}
and

{\footnotesize
\begin{equation}
\label{madash}
M_a^\prime=\Big\{2^{x-1}+\sum_{j=0}^{d-1}x_j2^{2^j-1} ~\Big\vert ~ x=\sum_{j=0}^{d-1} x_j2^j \in M_a,x_j\in \mathbb{F}_2  \Big\}.
\end{equation}}

Note that the number of elements in $M_a$ is $a-d$. Moreover, $M_a\subseteq M_b$ and $M_a^\prime\subseteq M_b^\prime$ whenever $a\leq b$. When $a$ is a power of $2$, $M_a=M_{a-1}$ and $M_a^\prime=M_{a-1}^\prime$. \\
\begin{exmp}
The sets $M_a$  and $M_a^\prime$ for $a=3$ to $9$ are shown in Table \ref{tab1} at the top of this page.

\end{exmp}

\begin{lemma}
\label{pdist3}
 Let $M_a^\prime$ be as defined in \eqref{madash}. View $M_a^\prime$ as a subset of $V=\f_2^a$ by identifying each element of $M_a^\prime$ with its radix-2 representation vector of length $a$. Then the MHD of the linear space spanned by $M^\prime_a$, denoted by $S$, is $3$.
\end{lemma}
\begin{IEEEproof}
The subspace $S\subset V$ is given by \\
 $S=\{\sum_{j=0}^{a-d-1}c_jy_j^\prime ~\vert ~ y_j^\prime\in M_a^\prime \}$ where $c_j\in\f_2$ for $j=0,1,\cdots,{a-d-1}$.\\ 
Observe that the map given by
\begin{eqnarray}
\label{xdash}
\begin{array}{l}
 \hspace{40pt}      f : M_a \rightarrow M_a^\prime   \\
             x=\sum_{j=0}^{d-1} x_j2^j  \mapsto      x^\prime=2^{x-1}+\sum_{j=0}^{d-1}x_j2^{2^j-1},
\end{array}
\end{eqnarray}
is one-one. Thus, the size of $M_a^\prime$, denoted as $\lvert{M_a}^\prime\rvert$ is also $a-d$.
Notice that $2^{x-1}\neq 2^{2^j-1}$ for $j=0,1,\cdots,d-1$ as $x\neq 2^j$.
So, $wt(x^\prime)=1+ wt (x)$ for all $x\in M_a$ where $wt(x)$ stands for the Hamming weight of $x$.
Now $wt(x)\geq 2$ as $x$ is not a power of $2$. So $wt(x^\prime)\geq 3$.

Similarly, $wt(x^\prime\oplus y^\prime )=2+ wt (x\oplus y)$ for all $x,y\in M_a, x\neq y$.
Now $wt(x\oplus y)\geq 1$ as $x\neq y$, which implies that $wt(x^\prime\oplus y^\prime )\geq 3$ for all $x^\prime,y^\prime\in M_a^\prime$.

 In general, $wt(y_1^\prime \oplus y_2^\prime\oplus\cdots \oplus y_k^\prime)= k+ wt (y_1 \oplus y_2\oplus \cdots \oplus y_k )$ for $k\leq a-d$, $y_1^\prime \neq y_2^\prime\neq \cdots \neq y_k^\prime$.
So for all $k\geq 3$ and $k\leq a-d$, $wt(y_1^\prime \oplus y_2^\prime\oplus\cdots \oplus y_k^\prime)\geq 3$.
 Now there exists an element in $S$, for instance $7$, whose Hamming weight is $3$. Thus, the MHD of $S$ is $3$.
\end{IEEEproof}
\begin{lemma}
\label{part}
Let $a$ and $d$ be non-zero positive integers such that $2^{d-1} \leq a < 2^d$ and  $V_a=\{0,1,\cdots,2^a-1\}$.
Then, there exists a partition of  $V_a$ into $2^d$ subsets $C_j^{(a)}, j=0,1,\cdots,2^d-1$ each containing $2^{a-d}$ elements, such that for any two distinct elements $x,y\in C_j^{(a)}$, $j\in\{0,1,\cdots,2^d-1\}$, we have $N_x^{(a)}\cap N_y^{(a)} =\phi.$
\end{lemma}

\begin{IEEEproof}
We identify the set $V_a$ with $\f_2^a$ by viewing each element of $V_a$ with its radix-2 representation vector. Let $M_a^\prime$ be as given by \eqref{madash}
and  $S$ be the sub-space of $V_a$ spanned by the radix-2 representation vectors (of length $a$) of the elements of the set $M_a^\prime$. The number of elements in $S$ is $2^{a-d}$. By Lemma~\ref{pdist3}, the MHD  of $S$ is $3$. Now we define a relation $'\sim'$ on $V_a$ as follows:
For all $a,b\in V_a$, $a\sim b$, if $a\oplus b\in S$.
We observe that this is an equivalence relation as
 for all $a,b$ and $c\in V_a$,\\
\noindent
$1)~ a\sim a\mbox { as } a\oplus a=0\in S$,\\
$2)~ a\sim b \Rightarrow b\sim a \mbox { as } a\oplus b\in S, \mbox { implies that } b\oplus a\in S$.\\
$3)~ a\sim b \mbox{ and } b\sim c, \mbox{ then } a\sim c, \mbox{ as } a\oplus b\in S$ and $b\oplus c\in S, \mbox { together imply } a\oplus c\in S$.\\
The number of equivalence classes  is $\frac{2^a}{2^{a-d}}=2^d$ and these equivalence classes are denoted as $C_i^{(a)}, i=0,1,\cdots,2^d-1$.
For any one equivalence class $C_i^{(a)}$, the elements in $C_i^{(a)}$ are given by $\{x\oplus s\vert s\in S\}$ for some $x\in C_i^{(a)}$.
Now the MHD of the class $C_i^{(a)}$, is also equal to the MHD of $S$ which is 3.
By lemma~\ref{dist3}, $N_x^{(a)}\cap N_y^{(a)} =\phi$ for all $x,y\in C_i^{(a)}$,
$i=0$ to $2^d-1$.
\end{IEEEproof}


The following example illustrates the partition of $V_a$ into the subsets $C_i^{(a)}$, $i=0$ to $2^d-1$, for $a=3,4,5$ and $6$.
\begin{exmp}
(i) Let $a=3$. $V_3$ is partitioned into $4$ classes $C_0^{(3)},C_1^{(3)},C_2^{(3)}$ and $C_3^{(3)}$, each containing $2$ elements. We have already seen that $M_3=\{3\}$ and $M_3^\prime=\{7\}$. $C_i^{(a)}=\{i,i\oplus 7\}$ for $i=0,1,2$ and $3$. Explicitly, \\
$C_0^{(3)}=\{0,7\},C_1^{(3)}=\{1,6\}, C_2^{(3)}=\{2,5\}, C_3^{(3)}=\{3,4\}$. 

(ii) For $a=4$, \\
$C_0^{(4)}=\{0,7\},C_1^{(4)}=\{1,6\},C_2^{(4)}=\{2,5\},\\
 C_3^{(4)}=\{3,4\},C_4^{(4)}=\{8,15\},C_5^{(4)}=\{9,14\},\\
C_6^{(4)}=\{10,13\},C_7^{(4)}=\{11,12\}$.

(iii) For $a=5$,\\
$C_0^{(5)}=\{0,7,25,30\},C_1^{(5)}=\{1,6,24,31\},\\
C_2^{(5)}=\{2,5,27,28\}, C_3^{(5)}=\{3,4,26,29\},\\
C_4^{(5)}=\{8,15,17,22\},C_5^{(5)}=\{9,14,16,23\},\\
C_6^{(5)}=\{10,13,19,20\}, C_7^{(5)}=\{11,12,18,21\}$.

(iv) For $a=6$,\\
$C_0^{(6)}=\{0,7,25,30,42,45,51,52\}$, \\ $C_1^{(6)}=\{1,6,24,31,43,44,50,53\}$,\\
$C_2^{(6)}=\{2,5,27,28,40,47,49,54\}$, \\  $C_3^{(6)}=\{3,4,26,29,41,46,48,55\}$,\\
$C_4^{(6)}=\{8,15,17,22,34,37,59,60\}$, \\ $C_5^{(6)}=\{9,14,16,23,35,36,58,61\}$,\\
$C_6^{(6)}=\{10,13,19,20,32,39,57,62\}$, \\  $C_7^{(6)}=\{11,12,18,21,33,38,56,63\}$.\\

\end{exmp}


\begin{theorem}
\label{fracz}
Let $a$ and $d$ be non-zero positive integers and $2^{d-1} \leq a < 2^d$.
There exists a SCOD $\mathbf{H}_a$ of size $2^a\times 2^a$ with entries of the matrix $2^{\frac{a-d}{2}}\mathbf{H}_a$ consisting $\pm x_1, \pm x_2,\cdots,\pm x_{a+1}$ or their conjugates,
such that the code has rate $\mathcal{R}=\frac{a+1}{2^a}$ and the ratio of number of zeros to the total number of entries of the matrix is equal to  $1-\mathcal{R}\cdot 2^{\lfloor\log_2{\frac{1}{\mathcal{R}}}\rfloor}$.
\end{theorem}
\begin{IEEEproof}
The SCOD $\mathbf{H}_a$ satisfying the required rate and fraction of zeros in the matrix is obtained from the COD $\mathbf{G}_a$ of size $2^a\times 2^a$ (given in \eqref{itcod}) as follows:

The rate $\mathcal{R}$ of the COD $\mathbf{G}_a$ is $\frac{a+1}{2^a}$.
Using Lemma~\ref{part},  $2^a$ rows of the COD $\mathbf{G}_a$ can be partitioned into $2^d$ groups with each group containing $2^{a-d}$ rows such that  any two distinct rows from any of $2^d$ groups, is non-intersecting.
If we add or subtract all the rows in a given class, the resulting row will not have any entry which is a linear combination of two or more variables.
Labeling the groups as $C_0^{(a)},C_1^{(a)},\cdots,C_{2^d-1}^{(a)}$, we define the $2^{a-d}\times 2^a$ matrices $\mathbf{B}_i$  formed by the rows of $\mathbf{G}_a$ which are in $C_i^{(a)}$ for $i=0$ to $2^d-1$.
Now form the matrix 
\begin{equation}
\mathbf{B}^\prime=
  \left[\begin{array}{c}
  \mathbf{B}_0 \\
  \mathbf{B}_1 \\
  \vdots \\
  \mathbf{B}_{2^d-1}
 \end{array}\right].
\end{equation}
The matrix $\mathbf{B}^\prime$ is of size $2^a\times 2^a$ and it is related to $\mathbf{G}_a$ by $\mathbf{B}^\prime=\mathbf{P}\mathbf{G}_a$
where $\mathbf{P}$ is a permutation matrix of size $2^a\times 2^a$.
We consider a Hadamard matrix $\mathbf{H}$ of size $2^{a-d}\times 2^{a-d}$ containing $1$ and $-1$ such that $\mathbf{H}^\mathcal{T}\mathbf{H}= 2^{a-d}\mathbf{I}_{2^{a-d}}$. Let $\widetilde{\mathbf{B}}_i=\mathbf{H}\mathbf{B}_i$.
The required matrix $\mathbf{H}_a$ is
\begin{eqnarray}
\mathbf{H}_a=2^{-\frac{a-d}{2}}
  \left[\begin{array}{c}
  \widetilde{\mathbf{B}}_0 \\
  \widetilde{\mathbf{B}}_1 \\
   \vdots \\
  \widetilde{\mathbf{B}}_{2^d-1}
 \end{array}\right].
\end{eqnarray}
For two matrices $\mathbf{A}=[a_{ij}]$ and $\mathbf{B}$, the tensor product of $\mathbf{A}$ with $\mathbf{B}$, denoted by $\mathbf{A}\otimes \mathbf{B}$, is the matrix
 $[a_{i,j}\mathbf{B}]$.
Let $\widetilde{\mathbf{H}}=\mathbf{I}_{2^d}\otimes \mathbf{H}$. 
We can write $\mathbf{H}_a=2^{-\frac{a-d}{2}}\widetilde{\mathbf{H}}\mathbf{B}^\prime=2^{-\frac{a-d}{2}}\widetilde{\mathbf{H}}\mathbf{P}\mathbf{G}_a$.
Now $\mathbf{H}_a$ is a SCOD if and only if $2^{-\frac{a-d}{2}}\widetilde{\mathbf{H}}\mathbf{P}$ is an unitary matrix.
As $\mathbf{P}$ is permutation matrix, it is enough to prove that $2^{-\frac{a-d}{2}}\widetilde{\mathbf{H}}$ is an unitary matrix.
Indeed, $\widetilde{\mathbf{H}}^\mathcal{T}\widetilde{\mathbf{H}}=\mathbf{I}_{2^d}\otimes (\mathbf{H}^\mathcal{T}\mathbf{H})=2^{a-d}\mathbf{I}_{2^a}$, thus $\mathbf{H}_a$ is a SCOD.

The number of locations containing $0$ in any row of $\mathbf{H}_a$ is $2^a-(a+1)2^{a-d}$.
Hence the fraction of zeros in $\mathbf{H}_a$ is equal to $\frac{2^a-(a+1)2^{a-d}}{2^a} = 1-\frac{a+1}{2^a}2^{a-d}$.
Now $2^{a-d}\leq\frac{2^a}{a+1}< 2^{a-d+1}$ as $2^d\geq {a+1}>2^{d-1}$. So, $a-d\leq\log_2{\frac{2^a}{a+1}}< a-d+1$ and  $a-d =\lfloor\log_2{\frac{2^a}{a+1}}\rfloor = \lfloor\log_2{\frac{1}{\mathcal{R}}}\rfloor$.\\
Thus the fraction of zeros is $1-\mathcal{R}\cdot 2^{\lfloor\log_2{\frac{1}{\mathcal{R}}}\rfloor}$.
\end{IEEEproof}

The proof of Theorem \ref{fracz} suggests a  recipe to construct the SCOD $\mathbf{H}_a$ with the fraction of zeros specified in the statement, from a COD $\mathbf{G}_a$ given in \eqref{itcod}. The following example illustrates this recipe.
\begin{exmp}
We consider the construction of rate-$1/2$, $8\times 8$ COD with no zero in the matrix. Following the recipe described above, the  permutation matrix $\mathbf{P}$ and $\widetilde{\mathbf{H}}$ are given by

{\footnotesize
\begin{eqnarray*}
 \mathbf{P}=\left[\begin{array}
{r@{\hspace{0.2pt}}r@{\hspace{0.2pt}}r@{\hspace{0.2pt}}r@{\hspace{0.2pt}}r@{\hspace{0.2pt}}r@{\hspace{0.2pt}}r@{\hspace{0.2pt}}r@{\hspace{0.2pt}}}
  1&0&0&0&0&0&0&0 \\
  0&0&0&0&0&0&0&1  \\
  0&1&0&0&0&0&0&0  \\
  0&0&0&0&0&0&1&0  \\
  0&0&1&0&0&0&0&0 \\
  0&0&0&0&0&1&0&0 \\
  0&0&0&1&0&0&0&0 \\
  0&0&0&0&1&0&0&0
\end{array}\right],
 \widetilde{\mathbf{H}}= \left[\begin{array}
{r@{\hspace{0.2pt}}r@{\hspace{0.2pt}}r@{\hspace{0.2pt}}r@{\hspace{0.2pt}}r@{\hspace{0.2pt}}r@{\hspace{0.2pt}}r@{\hspace{0.2pt}}r@{\hspace{0.2pt}}}
  1&1&0&0&0&0&0&0 \\
  1&-1&0&0&0&0&0&0  \\
  0&0&1&1&0&0&0&0  \\
  0&0&1&-1&0&0&0&0  \\
  0&0&0&0&1&1&0&0 \\
  0&0&0&0&1&-1&0&0 \\
  0&0&0&0&0&0&1&1 \\
  0&0&0&0&1&0&1&-1
\end{array}\right]
\end{eqnarray*}
}
respectively.
The matrix $\mathbf{H}_3=2^{-\frac{1}{2}}\widetilde{\mathbf{H}}\mathbf{P}\mathbf{G}_3$ is given by

{\small \begin{equation*}
\frac{1}{\sqrt{2}}\left[\begin{array}{rrrrrrrr}
   x_1   &-x_2^*  &-x_3^*& x_4  &-x_4^* &-x_3   & x_2  & x_1^* \\
   x_1   &-x_2^*  &-x_3^*&-x_4  &-x_4^* & x_3   &- x_2 &- x_1^*\\
   x_2   & x_1^*  & x_4  &-x_3^*&-x_3   &-x_4^* & x_1  &-x_2^*\\
   x_2   & x_1^*  &-x_4  &-x_3^*& x_3   &-x_4^* &-x_1  & x_2^*\\
   x_3   & x_4    & x_1^*& x_2^*&-x_2   & x_1   &-x_4^*& x_3^*\\
   x_3   &-x_4    & x_1^*& x_2^*& x_2   &-x_1   &-x_4^*&-x_3^*\\
   x_4   & x_3    &-x_2  & x_1  & x_1^* & x_2^* & x_3^*&-x_4^* \\
  -x_4   & x_3    &-x_2  & x_1  &- x_1^*&- x_2^*&-x_3^*&-x_4^*
\end{array}\right]
\end{equation*}
}
\noindent
which is a row permuted version of the code given in \eqref{matrix8}. 
\end{exmp}
\section{Premultiplying matrix}
\label{sec3}
In this section, we present a procedure to compute a matrix denoted as $\mathbf{Q}^{(a)}$ of size $2^a\times 2^a$ which when  pre-multiplies  $\mathbf{G}_a,$ along with an appropriate scaling factor, the resulting matrix $\mathbf{H}_a$ is the SCOD with desired fraction of zeros. The scaling factor, when multiplied to the matrix $\mathbf{Q}^{(a)}$, makes it a unitary matrix.

In order to construct the matrix $\mathbf{Q}^{(a)}$, we first associate a $2^x\times 2^x$ matrix $\mathbf{Q}_x$ to each $x$ in $M_a$.
The $(i,j)$th element of $\mathbf{Q}_x$, denoted by $\mathbf{Q}_x(i,j),$ is defined as follows:\\
For $i=0 \text { to }2^{x-1}-1,$
\begin{enumerate}
\item $\mathbf{Q}_x (i,i)=1;$
\item $\mathbf{Q}_x (i,i \oplus x^\prime)=1;$
\item $\mathbf{Q}_x (i \oplus x^\prime,i)=1;$
\item $\mathbf{Q}_x (i \oplus x^\prime,i \oplus x^\prime)=-1;$
\item $\mathbf{Q}_x(i,j)=0$ for all other values of $i$ and $j$;
\end{enumerate}
\noindent
where $x^\prime$ is given by \eqref{xdash}.
Define a $2^a\times 2^a$ matrix  $\widetilde{\mathbf{Q}}_x$ as $\widetilde{\mathbf{Q}}_x=\mathbf{I}_{2^{a-x}} \otimes \mathbf{Q}_x$ where $\mathbf{I}_{2^{a-x}}$ is the identity matrix of size $2^{a-x}\times 2^{a-x}$. \\
\begin{exmp}
In this example, we compute  $\mathbf{Q}_x$ for $x\in M_4$.
We have $M_4=\{3\}$. The matrix $\mathbf{Q}_3$ is the matrix shown on the left hand side of \eqref{matrix8} without the scaling factor $\frac{1}{\sqrt{2}}.$
\end{exmp}

Towards the construction of premultiplying matrix, we need the following result which says that $\widetilde{\mathbf{Q}}_x$ and $\widetilde{\mathbf{Q}}_y$ commute for all $x,y\in M_a$.
\begin{lemma}
\label{CommuteQ}
$\widetilde{\mathbf{Q}}_x\widetilde{\mathbf{Q}}_y=\widetilde{\mathbf{Q}}_y\widetilde
{\mathbf{Q}}_x$ for all $x,y\in M_a$.
\end{lemma}
\begin{IEEEproof}
Let the $i$th row of $\widetilde{\mathbf{Q}}_{x}$ be $\widetilde{\mathbf{Q}}_{x}^i$.
There are $2^a$ rows and $2^a$ columns for the matrix $\widetilde{\mathbf{Q}}_{x}$ with non-zero entries either $1$ or $-1$.\\ 
Fix $i$. Let $\widetilde{\mathbf{Q}}_{x}(i,l)=c_l$, $c_l\in \{0,1,-1\}$ for $l=0$ to $2^a-1$.
So $\widetilde{\mathbf{Q}}_{x}^i=(c_0,c_1,\cdots,c_{2^a-1})$.
We write $(c_0,c_1,\cdots,c_{2^a-1})$ as $\sum_{l=0}^{2^a-1}c_l2^l$ and this correspondence is unique in the sense that no two distinct vectors will produce the same value  under the above correspondence.
Then, we have $\widetilde{\mathbf{Q}}_{x}^i=\sum_{l=0}^{2^a-1}c_l2^l$.
Let the radix-2 representation of $i$ and $j$ be $(i_{a-1}, i_{a-2},\cdots, i_0)$ and $(j_{a-1}, j_{a-2},\cdots, j_0)$ respectively.\\
Note that $\widetilde{\mathbf{Q}}_{x}^i=(-1)^{i_{x-1}}2^i + 2^{i\oplus x^\prime}$ for $i=0$ to $2^a-1$.
Moreover, $\widetilde{\mathbf{Q}}_{x}$ is a symmetric matrix for all $x\in M_a$.
Let $K=\widetilde{\mathbf{Q}}_x\widetilde{\mathbf{Q}}_y$ and the $(i,j)$th entry of $\mathbf{K}$ be $\mathbf{K}(i,j)$.
It follows that
  \begin{eqnarray*}
\begin{array}{c}
    \mathbf{K}(i,j) =
    \begin{cases}
          0          & \text{ if $i\oplus j\notin \{0,x^\prime,y^\prime, x^\prime\oplus y^\prime\}$ }\\
      (-1)^{i_{x-1}} & \text{ if $i\oplus j= y^\prime$ }\\
      (-1)^{j_{y-1}} & \text{ if $i\oplus j= x^\prime$ }\\
         1           & \text{ if $i\oplus j= x^\prime \oplus y^\prime$ }\\
     (-1)^{i_{x-1}+ j_{y-1}} & \text{ if $i\oplus j= 0$ },\\
    \end{cases} \\ 
    \mathbf{K}(j,i) =
    \begin{cases}
          0          & \text{ if $i\oplus j\notin \{0,x^\prime,y^\prime,x^\prime \oplus y^\prime\}$ }\\
      (-1)^{j_{x-1}} & \text{ if $i\oplus j= y^\prime$ }\\
      (-1)^{i_{y-1}} & \text{ if $i\oplus j= x^\prime$ }\\
         1           & \text{ if $i\oplus j= x^\prime \oplus y^\prime$ }\\
     (-1)^{j_{x-1}+ i_{y-1}} & \text{ if $i\oplus j= 0$ }.\\
    \end{cases}\\
\end{array}
     \end{eqnarray*}

Let $x=\sum_{l=0}^{d-1}x_l2^l$ and $y=\sum_{l=0}^{d-1}y_l2^l$.
We have $y^\prime= 2^{y-1}+\sum_{l=0}^{d-1}y_l2^{2^l-1}$.
Note that the $(x-1)$th component of $y^\prime$, i.e., the coefficient of $2^{x-1}$ in the radix-2 representation of $y^\prime$ is zero as $x\neq y$ and $x$ is not a power of $2$.
Thus ${i_{x-1}}={j_{x-1}}$ when $i\oplus j= y^\prime$.
Similarly ${j_{y-1}}={i_{y-1}}$ when $i\oplus j= x^\prime$ and $i_{x-1}+ j_{y-1}=j_{x-1}+ i_{y-1}$ when $i\oplus j=0$, i.e., $i=j$.
Thus $\mathbf{K}(i,j)=\mathbf{K}(j,i)$ for all $i,j\in\{0,1,\cdots,2^a-1\}$.
So, $\mathbf{K}=\widetilde{\mathbf{Q}}_x\widetilde{\mathbf{Q}}_y$ is symmetric.
Then, $\mathbf{K}^\mathcal{T}={\widetilde{\mathbf{Q}}_y}^\mathcal{T}{\widetilde{\mathbf{Q}}_x}^\mathcal{T}=\mathbf{K}=\widetilde{\mathbf{Q}}_x\widetilde{\mathbf{Q}}_y$.
As $\widetilde{\mathbf{Q}}_y$ and $\widetilde{\mathbf{Q}}_x$ are symmetric matrix, they commute.
\end{IEEEproof}

Let $\mathbf{Q}^{(a)}=\underset{x\in M_a}{\prod}\widetilde{\mathbf{Q}}_x$, which is well defined since the product of these matrices does not depend on the order these matrices are multiplied. The following theorem asserts that $\mathbf{Q}^{(a)}$ so constructed, will produce a SCOD with the desired fraction of zeros.
\begin{theorem}
\label{premat}
Let $a$ and $d$ be non-zero positive integers and $2^{d-1} \leq a < 2^d$.
Then $\mathbf{H}_a=2^{-\frac{a-d}{2}}\mathbf{Q}^{(a)}\mathbf{G}_a$ is the desired SCOD with the rate and the fraction of zeros as specified in Theorem~\ref{fracz}.
\end{theorem}
\begin{IEEEproof}
We have to prove that $\mathbf{H}_a=2^{-\frac{a-d}{2}}\mathbf{Q}^{(a)}\mathbf{G}_a$ is a SCOD of size $2^a\times 2^a$ with rate $\mathcal{R}=\frac{a+1}{2^a}$ and the fraction of zeros is equal to  $1-\mathcal{R}\cdot 2^{\lfloor\log_2{\frac{1}{\mathcal{R}}}\rfloor}$.
Since we obtain $\mathbf{H}_a$ by pre-multiplying a constant matrix to $\mathbf{G}_a$, the rate remains same.
Thus, it is enough to prove that $\mathbf{H}_a$ is a SCOD and it contains the desired fraction of zeros.
As $\mathbf{G}_a$ is a COD, so $\mathbf{H}_a$ is a COD if $2^{-\frac{a-d}{2}}\mathbf{Q}^{(a)}$ is an unitary matrix.
Moreover, if each row of $\mathbf{H}_a$ contains $2^{a-d}(a+1)$ non-zero entries, 
then it contains the required fraction of zeros.
It is easy to note that if the matrix $\mathbf{Q}^{(a)}$ has $2^{a-d}$ non-zero elements in each of its rows, then each row of $\mathbf{H}_a$ will have $2^{a-d}(a+1)$ non-zero entries. The column co-ordinates of the non-zero entries of $i$th row of $\mathbf{Q}^{(a)}$ be such that the resulting matrix $\mathbf{H}_a$ will not have any entry which is linear combination of complex variables, i.e., those rows will be added or subtracted which possess non-intersecting property.
Thus we have to prove following two claims:
\begin{enumerate}
\item $2^{-\frac{a-d}{2}}\mathbf{Q}^{(a)}$ is an unitary matrix;
\item  The column co-ordinates of non-zero entries on the $i$-th row of $\mathbf{Q}^{(a)}$
    is given by the set $S_i^a=\{s\oplus i\vert s\in S^a\}$  where the subspace $S^a\subset \f_2^a$ is spanned by the set $M_a^\prime\subset \f_2^a$.
\end{enumerate}
We know that $\lvert{M_a}^\prime\rvert$ is $a-d$, thus $\lvert S^a\rvert$ is $2^{a-d}$.\\ First we prove claim 1).
Let $x\in M_a$.
Consider the $i$-th and $j$-th column of $\mathbf{Q}_x$ which are denoted as $\mathbf{Q}_x^i$ and $\mathbf{Q}_x^j$ respectively.
The inner product of $\mathbf{Q}_x^i$ and $\mathbf{Q}_x^j$, denoted as $\langle \mathbf{Q}_x^i, \mathbf{Q}_x^j\rangle$ is  as follows:
\begin{enumerate}
\item  for $j\neq i,i\oplus x^\prime$, $\langle \mathbf{Q}_x^i, \mathbf{Q}_x^j\rangle=0,$
\item  for $j=i\oplus x^\prime$, $\langle \mathbf{Q}_x^i, \mathbf{Q}_x^j\rangle= 1-1=0,$
\item  For $j=i$ ,$\langle \mathbf{Q}_x^i, \mathbf{Q}_x^j\rangle = 1+1 =2.$
 \end{enumerate}
So $\mathbf{Q}_x^\mathcal{T}\mathbf{Q}_x=2\mathbf{I}_{2^x}$ where $\mathbf{I}_{2^x}$ is an $2^x\times 2^x$ identity matrix.
Now $\widetilde{\mathbf{Q}}_{x}^\mathcal{T}\widetilde{\mathbf{Q}}_{x}=\mathbf{I}_{2^{a-x}} \otimes \mathbf{Q}_{x}^\mathcal{T}\mathbf{Q}_{x}$ which implies that
$\widetilde{\mathbf{Q}}_{x}^\mathcal{T}\widetilde{\mathbf{Q}}_{x}=2\mathbf{I}_{2^a}$
and ${\mathbf{Q}^{(a)}}^\mathcal{T}\mathbf{Q}^{(a)}=2^{a-d}\mathbf{I}_{2^a}$. Thus, $2^{-\frac{a-d}{2}}\mathbf{Q}^{(a)}$ is an unitary matrix.\\
Next, we prove claim 2) by induction on $a$.
Let $a=3$, then $\mathbf{Q}^{(3)}=\widetilde{\mathbf{Q}}_3=\mathbf{Q}_3$.
The $i$-th row of $\mathbf{Q}_3$ contains non-zero entries only at $i$ and $i\oplus 7$, corresponding to the elements of $\{s\oplus i\vert s\in S^3\}$ where $S^3=\{0,7\}$. Observe that $S^3$ is the subspace spanned by $M_3^\prime=\{7\}\subset \f_2^3$.\\
Let it be true for $a\leq (n-1)$.
Under this assumption, the column co-ordinate of the non-zero entries on the $i$th row of the matrix $\mathbf{Q}^{(n-1)}$ is given by the set $S_i^{n-1}=\{s\oplus i\vert s\in S^{n-1}\}$ for $i=0$ to $2^{n-1}-1$.
We have the following two cases:\\
{\bf Case (i)  $n$ is a power of $2$:}
In this case $M_n^\prime=M_{n-1}^\prime$ and $\mathbf{Q}^{(n)}=(\mathbf{I}_{2}\otimes \mathbf{Q}^{(n-1)})$. fact 2) is trivially satisfied. 
\noindent
{\bf Case (ii)  $n$ is not a power of $2$:}
$M_n=M_{n-1}\cup \{n\}$ and
$\mathbf{Q}^{(n)}=\widetilde{\mathbf{Q}}_{n}(\mathbf{I}_{2}\otimes \mathbf{Q}^{(n-1)})=\mathbf{Q}_{n}(\mathbf{I}_{2}\otimes \mathbf{Q}^{(n-1)})$.
The $i$-th row of ${\mathbf{Q}}_n$ is given by  $\widetilde{\mathbf{Q}}_{n}^i=(-1)^{i_{n-1}}2^i + 2^{i\oplus n^\prime}$ for $i=0$ to $2^n-1$ and
$j$-th row of $\mathbf{Q}^{(n-1)}$, denoted by $\mathbf{Q}^{(n-1),j}$ is $\sum_{s\in S_j^{n-1}} (-1)^{\alpha_s}2^s$ where $\alpha_s$ is either $+1$ or $-1$, depending on $s$.
The $(i,j)$-th entry of $\mathbf{Q}^{(n)}$ is given by the inner product of the $i$-th row of ${\mathbf{Q}}_n$ with $j$ th row of $(\mathbf{I}_{2}\otimes \mathbf{Q}^{(n-1)})$ as $\mathbf{Q}^{(n-1)}$ and  $(\mathbf{I}_{2}\otimes \mathbf{Q}^{(n-1)})$ both are symmetric matrix.
Viewing $S_i^{n-1}\subset\f_2^{n-1}$, as a subset of $\f_2^n$, by identifying $\f_2^{n-1}$ as a subspace of $\f_2^n$, it is possible to express $S_i^n$ in term of $S_i^{n-1}$.
The column co-ordinate of the non-zero entries at the $i$-th row of
$\mathbf{Q}^{(n)}$ is $S_i^n=S_i^{n-1}\cup \{s\oplus n^\prime\vert s\in S_i^{n-1}\}$ for $i=0$ to
$2^n-1$. Thus $S_i^n= \{s\oplus i\vert s\in S^n\}$.
\end{IEEEproof}

The following theorem shows that the matrix $\mathbf{H}_a$ with required rate and the fraction of zeros as specified in Theorem~\ref{fracz}, can also be constructed recursively, in a similar fashion as for $\mathbf{G}_a,$ and it is done using the premultiplying matrix $\mathbf{Q}^{(a)}.$ 

\begin{theorem}
Let $a$ be a non-zero positive integer and $\mathbf{H}_a$ be the SCOD as stated in Theorem~\ref{premat}. Then $\mathbf{H}_{a+1}$ is constructed recursively using $\mathbf{H}_a$ as follows:
\begin{eqnarray*}
\mathbf{H}_{a+1}=
\begin{cases}
 \mathbf{H}_{a+1}^\prime & \text { when  $(a+1)$ is a power of $2$ }\\
\frac{1}{\sqrt{2}}\mathbf{Q}_{a+1}\mathbf{H}_{a+1}^\prime & \text { otherwise },
\end{cases}
\end{eqnarray*}
where
\begin{eqnarray*}
     \mathbf{H}_{a+1}^\prime=\left[\begin{array}{rr}
 \mathbf{H}_a(x_1,x_2,\cdots,x_{a+1}) & -x_{a+2}^*\mathbf{H}_a(1,0,\cdots,0)\\
  x_{a+2} \mathbf{H}_a(1,0,\cdots,0)   &  \mathbf{H}_a(x_1^*,-x_2,\cdots,-x_{a+1})
\end{array}\right].
\end{eqnarray*}
\end{theorem}
\begin{IEEEproof} We have
\begin{eqnarray*}
M_{a+1}=
\begin{cases}
M_a & \text { when  $(a+1)$ is a power of $2$ }, \\
M_a\cup\{a+1\} & \text { otherwise }.
\end{cases}
\end{eqnarray*}
Now $\mathbf{Q}^{(a)}=\underset{x\in M_a}{\prod}\widetilde{\mathbf{Q}}_x$ where $\widetilde{\mathbf{Q}}_x$ is $2^a\times 2^a$ matrix.\\
To indicate the dependence of the size of $\widetilde{\mathbf{Q}}_{x}$ on $a$, we write $\widetilde{\mathbf{Q}}_{x}$ as $\widetilde{\mathbf{Q}}_{x}^a$. \\
We have $\widetilde{\mathbf{Q}}_x^a=\mathbf{I}_{2^{a-x}} \otimes \mathbf{Q}_x$ and $\widetilde{\mathbf{Q}}_x^{a+1}=\mathbf{I}_{2^{a+1-x}} \otimes \mathbf{Q}_x$.
So, $\widetilde{\mathbf{Q}}_x^{a+1}=\mathbf{I}_2\otimes \widetilde{\mathbf{Q}}_{x}^a$.\\
Moreover, $\mathbf{Q}^{(a)}=\underset{x\in M_a}{\prod}\widetilde{\mathbf{Q}}_x^a$ and $\mathbf{Q}^{(a+1)}=\underset{x\in M_{a+1}}{\prod}\widetilde{\mathbf{Q}}_x^{a+1}$.\\
Observe that, if $a+1$ is not a power of 2, then $$\mathbf{Q}^{(a+1)}=\widetilde{\mathbf{Q}}_{a+1}^{a+1}\cdot(\mathbf{I}_2\otimes \mathbf{Q}^{(a)}).$$\\ 
Since $\widetilde{\mathbf{Q}}_{a+1}^{a+1}=\mathbf{Q}_{a+1}$, we have 
\begin{eqnarray*}
\mathbf{Q}^{(a+1)}=
\begin{cases}
\mathbf{I}_2\otimes \mathbf{Q}^{(a)} & \text { when  $(a+1)$ is a power of $2$ }, \\
\mathbf{Q}_{a+1}\cdot(\mathbf{I}_2\otimes \mathbf{Q}^{(a)}) & \text { otherwise }.
\end{cases}
\end{eqnarray*}

From Theorem~\ref{premat}, we have $\mathbf{H}_a=2^{-\frac{a-d}{2}}\mathbf{Q}^{(a)}\mathbf{G}_a$ where $d$ is given by $2^{d-1} \leq a < 2^d$. Hence,

{\small
\begin{eqnarray*}
\mathbf{H}_{a+1}=
\begin{cases}
2^{-\frac{a-d}{2}}\cdot (\mathbf{I}_2\otimes \mathbf{Q}^{(a)})\cdot \mathbf{G}_{a+1} & \hspace*{-1.4cm}\text { when  $(a+1)$ is  a power of $2$, } \\
 2^{-\frac{a-d+1}{2}}\cdot \mathbf{Q}_{a+1}\cdot(\mathbf{I}_2\otimes \mathbf{Q}^{(a)})\cdot \mathbf{G}_{a+1} & \text { otherwise. }
\end{cases}
\end{eqnarray*}
}
Let
\begin{equation}
\label{recur1}
\mathbf{H}_{a+1}^\prime=2^{-\frac{a-d}{2}}(\mathbf{I}_2\otimes \mathbf{Q}^{(a)})\mathbf{G}_{a+1}.
\end{equation}
Then,
\begin{eqnarray*}
\mathbf{H}_{a+1}=
\begin{cases}
 \mathbf{H}_{a+1}^\prime & \text { when  $(a+1)$ is a power of $2$ },\\
\frac{1}{\sqrt{2}}\mathbf{Q}_{a+1}\mathbf{H}_{a+1}^\prime & \text { otherwise }.
\end{cases}
\end{eqnarray*}
 $\mathbf{H}_{a+1}^\prime$ is constructed using $\mathbf{H}_a$ as follows:\\
We have from \eqref{itcod},
\begin{equation}
\label{recur2}
\mathbf{G}_{a+1}=  \left[\begin{array}{rr}
\mathbf{G}_a   & -x_{a+2}^*\mathbf{I}_{2^a}      \\
x_{a+2}\mathbf{I}_{2^a}   & \mathbf{G}_a^{\mathcal{H}}
\end{array}\right].
\end{equation}

From the construction of $\mathbf{G}_a$ and $\mathbf{H}_a$, it follows that 
 \begin{eqnarray}
\label{recur3}
 \mathbf{I}_{2^a}=\mathbf{G}_a(1,0,\cdots,0),\mathbf{Q}^{(a)}=2^{\frac{a-d}{2}}\mathbf{H}_a(1,0,\cdots,0),\nonumber\\
\mathbf{G}_a^{\mathcal{H}}=\mathbf{G}_a^{\mathcal{H}}(x_1,x_2,\cdots,x_{a+1})=\mathbf{G}_a(x_1^*,-x_2,\cdots,-x_{a+1}).
 \end{eqnarray}
Using \eqref{recur1}, \eqref{recur2} and \eqref{recur3}, we have 

{\small
\begin{eqnarray*}
     \label{eqn_Hre}
     \mathbf{H}_{a+1}^\prime&=&\mathbf{H}_{a+1}^\prime(x_1,x_2,\cdots,x_{a+2})\\
         &=&\left[\begin{array}{rr}
 \mathbf{H}_a(x_1,x_2,\cdots,x_{a+1}) & -x_{a+2}^*\mathbf{H}_a(1,0,\cdots,0)\\
  x_{a+2} \mathbf{H}_a(1,0,\cdots,0)   &  \mathbf{H}_a(x_1^*,-x_2,\cdots,-x_{a+1})
\end{array}\right].
\end{eqnarray*}
}
\end{IEEEproof}
\begin{table*}
\caption{Comparison of power distribution characteristics}
\begin{center}
\begin{tabular}{|c||c||c||c||c|}\hline\label{tab2}
   & 16 Tx; QPSK & 16 Tx; 16 QAM & 32 Tx; QPSK & 32 Tx; 16 QAM \\
\begin{tabular}{c}
\\
\hline
SCODs $\mathbf{G}_a$ \\
\hline
Codes in this paper
\end{tabular}
&
\begin{tabular}{c|c}
Peak/ave & $P_0$ \\
\hline
3.2 & 0.6875 \\
\hline
1.6 & 0.375  \\
\end{tabular}
&
\begin{tabular}{c|c}
Peak/ave & $P_0$ \\
\hline
 $11.52$ & $0.6875$ \\
\hline
 $5.76$ & $0.375$ \\
\end{tabular}
&
\begin{tabular}{c|c}
Peak/ave & $P_0$ \\
\hline
 $5.33$  & $0.8125$\\
\hline
 $1.33$ & $0.25$ \\
\end{tabular}
&
\begin{tabular}{c|c}
Peak/ave & $P_0$ \\
\hline
$19.2$  & $0.8125$\\
\hline
$4.8$ & $0.25$\\
\end{tabular}
\\
\hline
\end{tabular}
\end{center}
\end{table*}

\begin{table*}
\caption{Variation of fraction of zeros with the number of antennas}
\begin{center}
\begin{tabular}{|l|l|l|l|l|l|l|l|l|l|l|l|l|l|l|} \hline  \label{tab3}
$a$ & 3 & 4& 5& 6& 7& 8  & 9 & 10 &11&12&13&14&15&16  \\ \hline 
$f_z(\mathbf{H}_a)$ & 0 & $\frac{3}{8}$& $\frac{2}{8}$&$\frac{1}{8}$ &0&$\frac{7}{16}$ &$\frac{6}{16}$ &$\frac{5}{16}$ &$\frac{4}{16}$&$\frac{3}{16}$&$\frac{2}{16}$&$\frac{1}{16}$ & 0 &$\frac{15}{32}$ \\ \hline  
$f_z(\mathbf{G}_a)$ &$1/2$ & $\frac{11}{16}$& $\frac{13}{16}$&$\frac{57}{64}$ &$\frac{120}{128}$&$\frac{247}{256}$ &$\frac{502}{512}$ &$\frac{1013}{1024}$&$\frac{2036}{2048}$&$\frac{4083}{4096}$&$\frac{8178}{8192}$
&$\frac{16369}{16384}$&$\frac{32752}{32768}$&$\frac{65519}{65536}$\\ \hline  \end{tabular}
\end{center}
\end{table*}

For $a=3$, $\mathbf{Q}^{(3)}$ is given in  \eqref{matrix8}. For $a=4$, $\mathbf{Q}^{(4)}$ is as follows:
{\footnotesize
\begin{equation*}
\label{cod9}
\left[\begin{array}{r @{\hspace{.4pt}}r @{\hspace{.4pt}}r @{\hspace{.4pt}}r @{\hspace{.4pt}}r @{\hspace{.4pt}}r @{\hspace{.4pt}}r @{\hspace{.4pt}}r @{\hspace{.4pt}}r @{\hspace{.4pt}}r @{\hspace{.4pt}}r @{\hspace{.4pt}}r @{\hspace{.4pt}}r @{\hspace{.4pt}}r @{\hspace{.4pt}}r @{\hspace{.4pt}}r @{\hspace{.4pt}}}
  1&0&0&0&0&0&0&1      &0&0&0&0&0&0&0&0\\
  0&1&0&0&0&0&1&0      &0&0&0&0&0&0&0&0 \\
  0&0&1&0&0&1&0&0      &0&0&0&0&0&0&0&0 \\
  0&0&0&1&1&0&0&0      &0&0&0&0&0&0&0&0 \\
  0&0&0&1&-&0&0&0     &0&0&0&0&0&0&0&0  \\
  0&0&1&0&0&-&0&0     &0&0&0&0&0&0&0&0  \\
  0&1&0&0&0&0&-&0     &0&0&0&0&0&0&0&0  \\
  1&0&0&0&0&0&0&-     &0&0&0&0&0&0&0&0\\
  0&0&0&0&0&0&0&0      &1&0&0&0&0&0&0&1 \\
  0&0&0&0&0&0&0&0      &0&1&0&0&0&0&1&0  \\
  0&0&0&0&0&0&0&0      &0&0&1&0&0&1&0&0  \\
  0&0&0&0&0&0&0&0      &0&0&0&1&1&0&0&0  \\
  0&0&0&0&0&0&0&0      &0&0&0&1&-&0&0&0  \\
  0&0&0&0&0&0&0&0      &0&0&1&0&0&-&0&0  \\
  0&0&0&0&0&0&0&0      &0&1&0&0&0&0&-&0  \\
  0&0&0&0&0&0&0&0      &1&0&0&0&0&0&0&-
\end{array}\right]
\end{equation*}
}
\noindent
The SCOD obtained by premultiplying $\mathbf{G}_4$ with $2^{-\frac{1}{2}}\mathbf{Q}^{(4)}$, is shown on the left hand side at the top of the next
page. For comparison, on the right hand side, we have displayed the SCOD $\mathbf{G}_4$ to compare the number of zeros. The premultiplying matrix $\mathbf{Q}^{(5)}$ for 32 antennas corresponding to $a=5$ is displayed in Fig. \ref{c1fig2} and the resulting code $\mathbf{H}_5$ in Fig. \ref{c1fig3}.
\begin{figure*}
{\tiny
\begin{eqnarray*}
\label{2in1}
\frac{1}{\sqrt{2}}
\left[
\begin{array}{r @{\hspace{.4pt}}r @{\hspace{.4pt}}r @{\hspace{.4pt}}r @{\hspace{.4pt}}r @{\hspace{.4pt}}r @{\hspace{.4pt}}r @{\hspace{.4pt}}r @{\hspace{.4pt}}r @{\hspace{.4pt}}r @{\hspace{.4pt}}r @{\hspace{.4pt}}r @{\hspace{.4pt}}r @{\hspace{.4pt}}r @{\hspace{.4pt}}r @{\hspace{.4pt}}r @{\hspace{.4pt}}}
    x_1  &-x_2^*  &-x_3^*& x_4  &-x_4^*&-x_3   & x_2  & x_1^*  &-x_5^* & 0 & 0 & 0 &0 & 0     & 0 & -x_5^*  \\ x_2  & x_1^*  & x_4  &-x_3^*&-x_3  &-x_4^* & x_1  &-x_2^*  &0 &-x_5^* & 0 & 0 &0 & 0 &-x_5^* & 0 \\
    x_3  & x_4    & x_1^*& x_2^*&-x_2  & x_1   &-x_4^*& x_3^*  &0 &0 &-x_5^* & 0 &0  & -x_5^* & 0 &0\\
    x_4  & x_3    &-x_2  & x_1  & x_1^*& x_2^* & x_3^*&-x_4^* &0 &0 &0  &-x_5^* & -x_5^* & 0 &0 &0 \\
   -x_4  & x_3    &-x_2  & x_1  &-x_1^*&-x_2^* &-x_3^*&-x_4^* &0 &0 &0  &-x_5^* & x_5^* & 0 &0 &0 \\
    x_3  &-x_4    & x_1^*& x_2^*& x_2  &-x_1   &-x_4^*&-x_3^*  &0 &0 &-x_5^* & 0 &0  & x_5^* & 0 &0\\
    x_2  & x_1^*  &-x_4  &-x_3^*& x_3  &-x_4^* &-x_1  & x_2^* &0 &-x_5^* & 0 & 0 &0 & 0 &x_5^* & 0\\
    x_1  &-x_2^*  &-x_3^*&-x_4  &-x_4^*&x_3    &-x_2  &-x_1^* &-x_5^* & 0 & 0 & 0 &0 & 0     & 0 & x_5^* \\
     x_5 & 0 & 0 & 0 &0 & 0     & 0 & x_5 &   x_1^*  & x_2^* & x_3^*&-x_4  & x_4^* &x_3    &-x_2  & x_1 \\
    0 & x_5 & 0 & 0 &0 & 0 & x_5 & 0 &-x_2   & x_1   &-x_4  & x_3^*& x_3   & x_4^* & x_1^*& x_2^* \\
    0 &0 &x_5 & 0 &0  & x_5 & 0 &0 &-x_3   &-x_4   & x_1  &-x_2^*& x_2   & x_1^* & x_4^*&-x_3^* \\
    0 &0 &0  & x_5 & x_5 & 0 &0 &0 &-x_4   &-x_3   & x_2  & x_1^*& x_1   &-x_2^* &-x_3^*& x_4^* \\
     0 &0 &0  & x_5 & -x_5 & 0 &0 &0 &x_4   &-x_3   & x_2  & x_1^*&-x_1   & x_2^* & x_3^*& x_4^* \\
    0 &0 & x_5 & 0 &0  & -x_5 & 0 &0 &-x_3   & x_4   & x_1  &-x_2^*&-x_2   &-x_1^* & x_4^*& x_3^*\\
    0 & x_5 & 0 & 0 &0 & 0 &-x_5 & 0 &-x_2   & x_1   &x_4   & x_3^*&-x_3   & x_4^* &-x_1^*&-x_2^*\\
      x_5 & 0 & 0 & 0 &0 & 0     & 0 &-x_5 &  x_1^* &x_2^*  & x_3^*& x_4  & x_4^* &-x_3   & x_2  &-x_1

\end{array}\right],
\left[
\begin{array}{r @{\hspace{.4pt}}r @{\hspace{.4pt}}r @{\hspace{.4pt}}r @{\hspace{.4pt}}r @{\hspace{.4pt}}r @{\hspace{.4pt}}r @{\hspace{.4pt}}r @{\hspace{.4pt}}r @{\hspace{.4pt}}r @{\hspace{.4pt}}r @{\hspace{.4pt}}r @{\hspace{.4pt}}r @{\hspace{.4pt}}r @{\hspace{.4pt}}r @{\hspace{.4pt}}r @{\hspace{.4pt}}}
x_1& -x_2^*& -x_3^*&  0& -x_4^*&  0&   0&      0& -x_5^*&  0&  0&  0&  0&  0&  0&  0\\
 x_2&x_1^*&  0& -x_3^*&  0& -x_4^*&  0&  0&  0& -x_5^*&  0&  0&  0&  0&  0&  0\\
 x_3&  0&x_1^*&x_2^*&  0&  0& -x_4^*&  0&  0&  0& -x_5^*&  0&  0&  0&  0&  0\\
         0& x_3&-x_2& x_1&  0&  0&  0& -x_4^*&  0&  0&  0& -x_5^*&  0&  0&  0&  0\\
 x_4&  0&  0&  0&x_1^*&x_2^*&x_3^*&  0&  0&  0&  0&  0& -x_5^*&  0&  0&  0\\
         0& x_4&  0&  0&-x_2& x_1&  0&x_3^*&  0&  0&  0&  0&  0& -x_5^*&  0&  0\\
         0&  0& x_4&  0&-x_3&  0& x_1& -x_2^*&  0&  0&  0&  0&  0&  0& -x_5^*&  0\\
         0&  0&  0& x_4&  0&-x_3& x_2&x_1^*&  0&  0&  0&  0&  0&  0&  0& -x_5^*\\
 x_5&  0&  0&  0&  0&  0&  0&  0&x_1^*&x_2^*&x_3^*&  0&x_4^*&  0&  0&  0\\
         0& x_5&  0&  0&  0&  0&  0&  0&-x_2& x_1&  0&x_3^*&  0&x_4^*&  0&  0\\
         0&  0& x_5&  0&  0&  0&  0&  0&-x_3&  0& x_1& -x_2^*&  0&  0&x_4^*&  0\\
         0&  0&  0& x_5&  0&  0&  0&  0&  0&-x_3& x_2&x_1^*&  0&  0&  0&x_4^*\\
         0&  0&  0&  0& x_5&  0&  0&  0&-x_4&  0&  0&  0& x_1& -x_2^*& -x_3^*&  0\\
         0&  0&  0&  0&  0& x_5&  0&  0&  0&-x_4&  0&  0& x_2&x_1^*&  0& -x_3^*\\
         0&  0&  0&  0&  0&  0& x_5&  0&  0&  0&-x_4&  0& x_3&  0&x_1^*&x_2^*\\
         0&  0&  0&  0&  0&  0&  0& x_5&  0&  0&  0&-x_4&  0& x_3&-x_2& x_1
\end{array}
\right]
\end{eqnarray*}
}
\hrule
\end{figure*}

\section{PAPR of the new codes}
\label{sec4}

In Table \ref{tab2}, the SCODs constructed in this paper are compared to the known SCODs given in \eqref{itcod}. For some fixed number of transmit antennas, we observe that the PAPR of these new codes is less compared to the known SCODs. In fact, it is easily seen  that as the number of transmit antenna increases, these codes outperform the existing CODs significantly as far as PAPR is concerned. Quantitatively, for QAM signal set, the PAPR of a SCOD for $2^a$ antennas given by \eqref{itcod}, is $\frac{2^a}{a+1}$, while it is $\frac{2^a}{(a+1)\cdot 2^{\lfloor log_2(\frac{2^a}{a+1}) \rfloor}}$ for the SCOD of same size constructed in this paper. The codes constructed in this paper contain fewer zeros than the well-known SCODs. Hence, the probability $P_0$ that an antenna transmits a zero symbol (or switched off) is less in these codes compared to the codes given in \eqref{itcod}. 

Another interesting fact to note is that the new SCODs for $2^a$ transmit antennas, contains no zero entry when $a+1$ is a power of $2$. For all other values of $a$, the fraction of zeros keeps reducing starting from $a=2^l$ to $2^{l+1}-1$, $l$ a positive integer. On the other hand, the fraction of zeros keeps increasing as $a$ ($\geq 3$) increases for the codes given in \eqref{itcod}. Table \ref{tab3} shows the variation in fraction of zeros (denoted as $f_z$) for proposed codes $\mathbf{H}_a$ and the SCODs $\mathbf{G}_a$ for $a=3$ to $16$.

\begin{figure*}[htp]
  \hfill
  \begin{minipage}[t]{.45\textwidth}
    \begin{center}  
      \epsfig{file=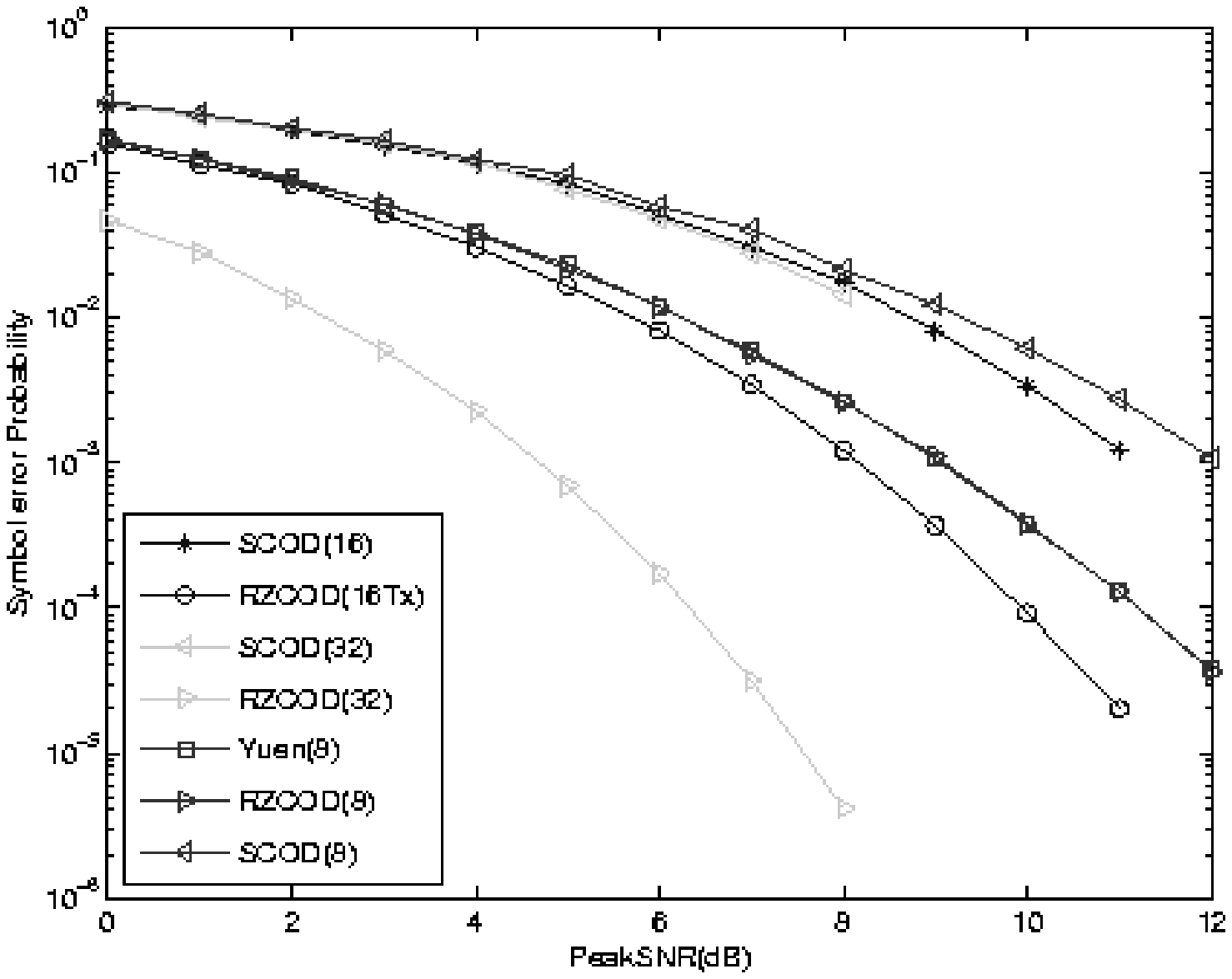, scale=0.5}
\caption{The performance of the RZCODs and SCODs for 8, 16 and 32 transmit antennas and the code given in Yuen et al for 8 transmit antennas using QAM modulation.} \label{fig1}
    \end{center}
  \end{minipage}
  \hfill
  \begin{minipage}[t]{.45\textwidth}
    \begin{center}  
      \epsfig{file=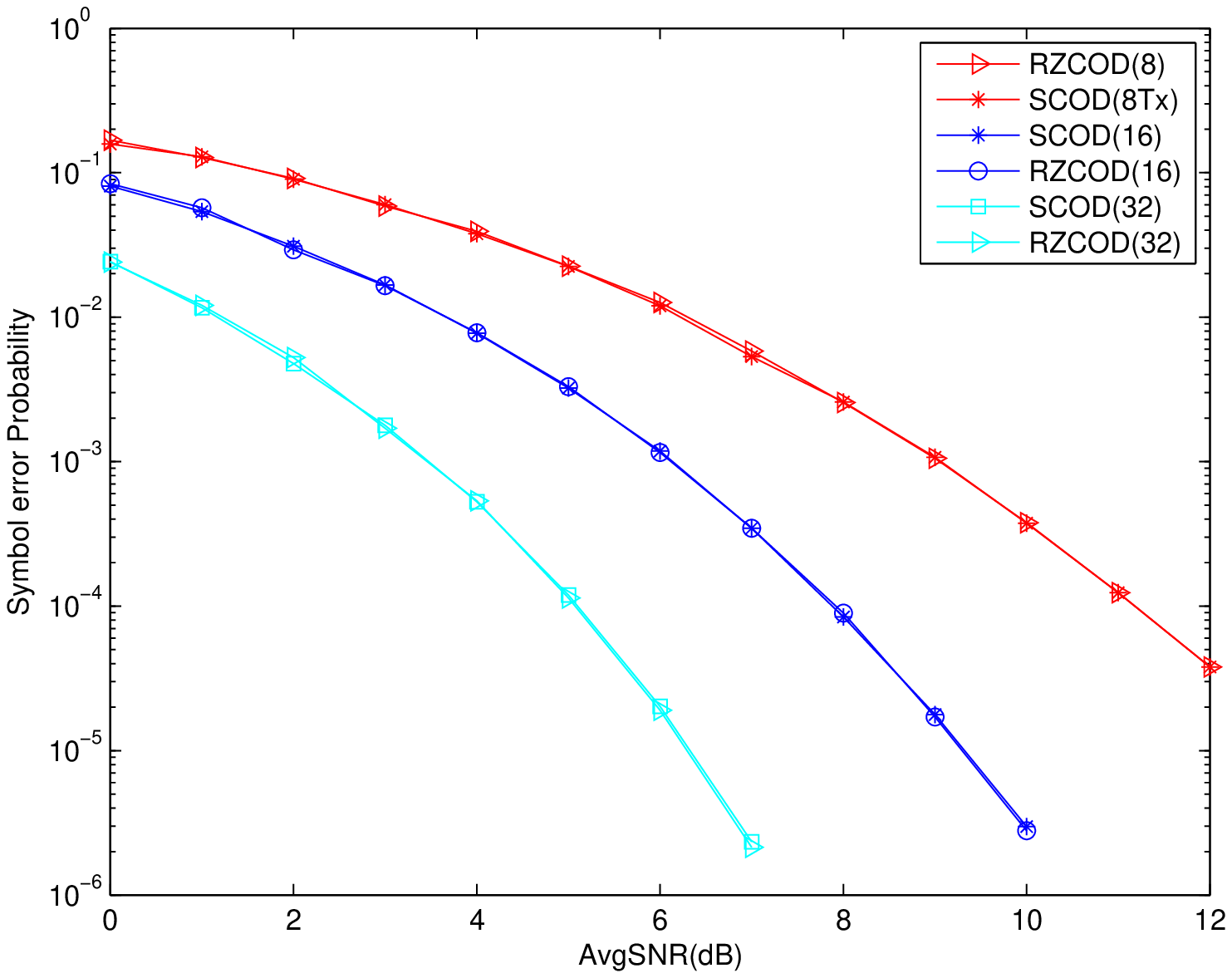, scale=0.5}
\caption{The performance of the RZCODs and SCODs for 8, 16 and 32 transmit antennas  using QAM modulation.} \label{figavg1}

    \end{center}
  \end{minipage}
  \hfill
  \hfill
  \begin{minipage}[t]{.45\textwidth}
    \begin{center}  
      \epsfig{file=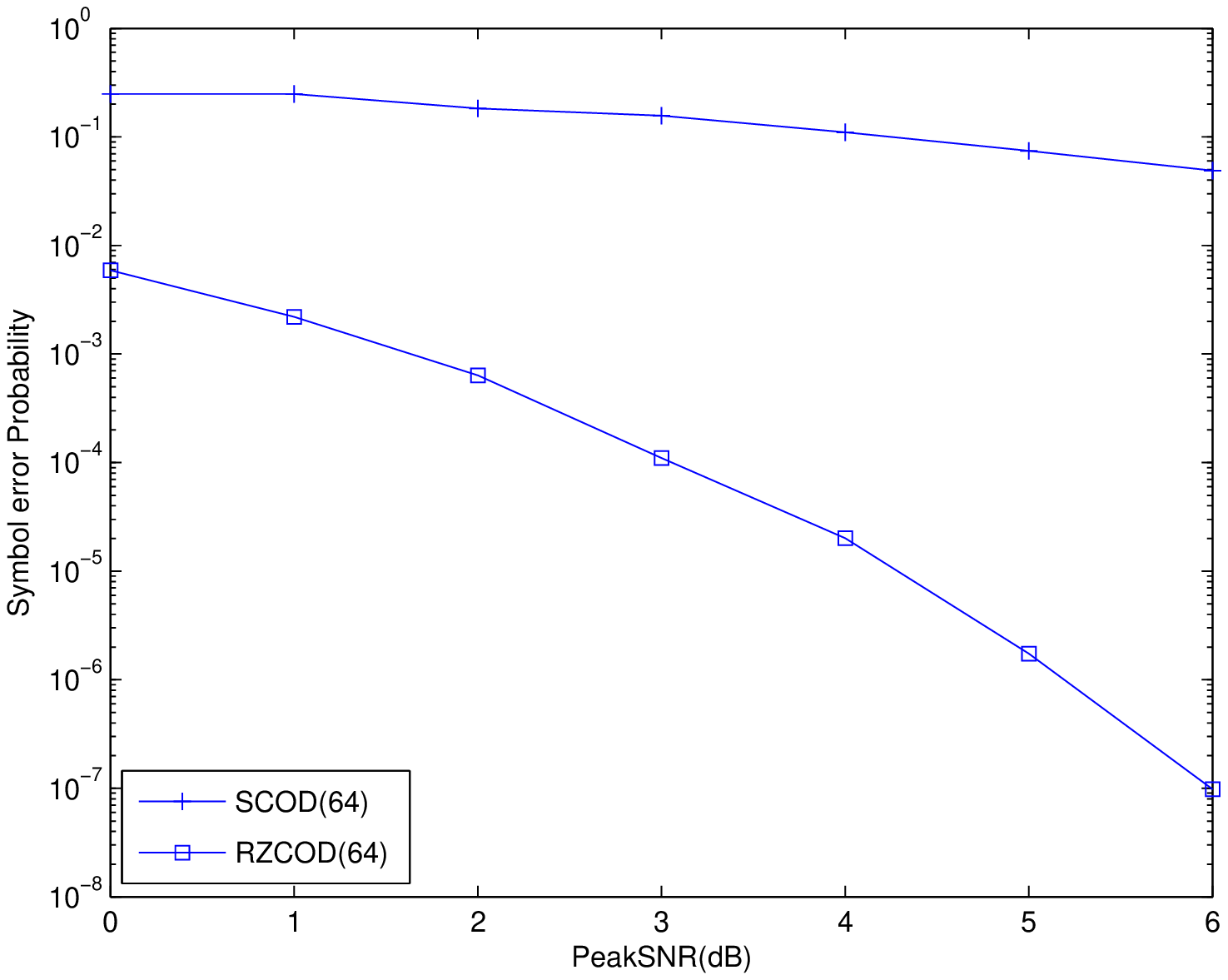, scale=0.5}
\caption{The performance of the RZCODs and SCODs for 64 transmit antennas using QAM modulation.} \label{fig2}
    \end{center}
  \end{minipage}
  \hfill
  \begin{minipage}[t]{.45\textwidth}
    \begin{center}  
      \epsfig{file=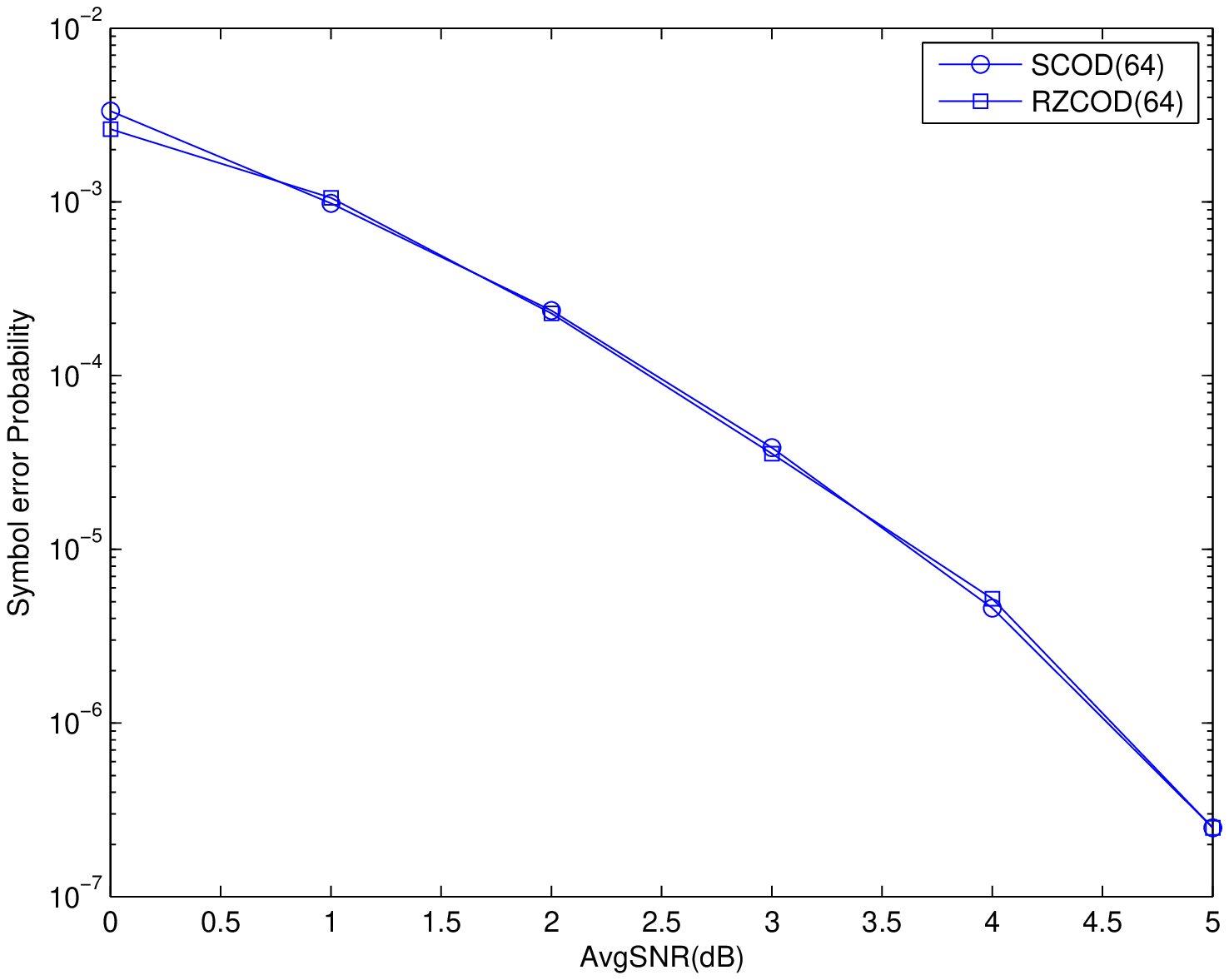, scale=0.5}
\caption{The performance of the RZCODs and SCODs for 64 transmit antennas using QAM modulation.} \label{figavg2}

    \end{center}
  \end{minipage}
  \hfill

\end{figure*}

\begin{figure*}

{\tiny
\begin{equation*}
\left[\begin{array}{cccccccccccccccccccccccccccccccc}
  1&0&0&0&0&0&0&1      &0&0&0&0&0&0&0&0  &0&0&0&0&0&0&0&0      &0&1&0&0&0&0&1&0\\
  0&1&0&0&0&0&1&0      &0&0&0&0&0&0&0&0  &0&0&0&0&0&0&0&0      &1&0&0&0&0&0&0&1\\
  0&0&1&0&0&1&0&0      &0&0&0&0&0&0&0&0  &0&0&0&0&0&0&0&0      &0&0&0&1&1&0&0&0  \\
  0&0&0&1&1&0&0&0      &0&0&0&0&0&0&0&0  &0&0&0&0&0&0&0&0      &0&0&1&0&0&1&0&0  \\
  0&0&0&1&-&0&0&0     &0&0&0&0&0&0&0&0  &0&0&0&0&0&0&0&0      &0&0&1&0&0&-&0&0 \\
  0&0&1&0&0&-&0&0     &0&0&0&0&0&0&0&0  &0&0&0&0&0&0&0&0      &0&0&0&1&-&0&0&0  \\
  0&1&0&0&0&0&-&0     &0&0&0&0&0&0&0&0  &0&0&0&0&0&0&0&0      &1&0&0&0&0&0&0&-  \\
  1&0&0&0&0&0&0&-     &0&0&0&0&0&0&0&0  &0&0&0&0&0&0&0&0      &0&1&0&0&0&0&-&0 \\
  0&0&0&0&0&0&0&0      &1&0&0&0&0&0&0&1  &0&1&0&0&0&0&1&0      &0&0&0&0&0&0&0&0\\
  0&0&0&0&0&0&0&0      &0&1&0&0&0&0&1&0  &1&0&0&0&0&0&0&1      &0&0&0&0&0&0&0&0\\
  0&0&0&0&0&0&0&0      &0&0&1&0&0&1&0&0  &0&0&0&1&1&0&0&0      &0&0&0&0&0&0&0&0\\
  0&0&0&0&0&0&0&0      &0&0&0&1&1&0&0&0  &0&0&1&0&0&1&0&0      &0&0&0&0&0&0&0&0\\
  0&0&0&0&0&0&0&0      &0&0&0&1&-&0&0&0 &0&0&1&0&0&-&0&0     &0&0&0&0&0&0&0&0 \\
  0&0&0&0&0&0&0&0      &0&0&1&0&0&-&0&0 &0&0&0&1&-&0&0&0     &0&0&0&0&0&0&0&0 \\
  0&0&0&0&0&0&0&0      &0&1&0&0&0&0&-&0 &1&0&0&0&0&0&0&-     &0&0&0&0&0&0&0&0\\
  0&0&0&0&0&0&0&0      &1&0&0&0&0&0&0&- &0&1&0&0&0&0&-&0     &0&0&0&0&0&0&0&0\\
  0&0&0&0&0&0&0&0      &0&1&0&0&0&0&1&0  &-&0&0&0&0&0&0&-    &0&0&0&0&0&0&0&0\\
  0&0&0&0&0&0&0&0      &1&0&0&0&0&0&0&1  &0&-&0&0&0&0&-&0    &0&0&0&0&0&0&0&0\\
  0&0&0&0&0&0&0&0      &0&0&0&1&1&0&0&0  &0&0&-&0&0&-&0&0    &0&0&0&0&0&0&0&0\\
  0&0&0&0&0&0&0&0      &0&0&1&0&0&1&0&0  &0&1&0&-&-&0&0&0    &0&0&0&0&0&0&0&0\\
  0&0&0&0&0&0&0&0      &0&0&1&0&0&-&0&0 &0&0&0&-&1&0&0&0     &0&0&0&0&0&0&0&0\\
  0&0&0&0&0&0&0&0      &0&0&0&1&-&0&0&0 &0&0&-&0&0&1&0&0     &0&0&0&0&0&0&0&0\\
  0&0&0&0&0&0&0&0      &1&0&0&0&0&0&0&- &0&-&0&0&0&0&1&0     &0&0&0&0&0&0&0&0\\
  0&0&0&0&0&0&0&0      &0&1&0&0&0&0&-&0 &-&0&0&0&0&0&0&1     &0&0&0&0&0&0&0&0\\
  0&1&0&0&0&0&1&0      &0&0&0&0&0&0&0&0  &0&0&0&0&0&0&0&0      &-&0&0&0&0&0&0&-\\
  1&0&0&0&0&0&0&1      &0&0&0&0&0&0&0&0  &0&0&0&0&0&0&0&0      &0&-&0&0&0&0&-&0\\
  0&0&0&1&1&0&0&0      &0&0&0&0&0&0&0&0  &0&0&0&0&0&0&0&0      &0&0&-&0&0&-&0&0\\
  0&0&1&0&0&1&0&0      &0&0&0&0&0&0&0&0  &0&0&0&0&0&0&0&0      &0&0&0&-&-&0&0&0\\
  0&0&1&0&0&-&0&0     &0&0&0&0&0&0&0&0  &0&0&0&0&0&0&0&0      &0&0&0&-&1&0&0&0\\
  0&0&0&1&-&0&0&0     &0&0&0&0&0&0&0&0  &0&0&0&0&0&0&0&0      &0&0&-&0&0&1&0&0\\
  1&0&0&0&0&0&0&-     &0&0&0&0&0&0&0&0  &0&0&0&0&0&0&0&0      &0&-&0&0&0&0&1&0\\
  0&1&0&0&0&0&-&0     &0&0&0&0&0&0&0&0  &0&0&0&0&0&0&0&0      &-&0&0&0&0&0&0&1
\end{array}
\right]
\end{equation*}
}
\caption{The premultiplying matrix $\mathbf{Q}^{(5)}$ for $32$ antennas}
\label{c1fig2}

%
%

{\tiny
\[
\hspace{-10pt}
\frac{1}{2}\left[ \hspace{-5pt}
\begin{array}{ r @{\hspace{.2pt}} r @{\hspace{.2pt}} r @{\hspace{.2pt}} r @{\hspace{.2pt}} r @{\hspace{.2pt}} r @{\hspace{.2pt}} r @{\hspace{.2pt}} r @{\hspace{.2pt}} r @{\hspace{.2pt}} r @{\hspace{.2pt}} r @{\hspace{.2pt}} r @{\hspace{.2pt}} r @{\hspace{.2pt}} r @{\hspace{.2pt}} r @{\hspace{.2pt}} r @{\hspace{.2pt}} r @{\hspace{.2pt}} r @{\hspace{.2pt}} r @{\hspace{.2pt}} r @{\hspace{.2pt}} r @{\hspace{.2pt}} r @{\hspace{.2pt}} r @{\hspace{.2pt}} r  @{\hspace{.2pt}}r @{\hspace{.2pt}} r @{\hspace{.2pt}} r @{\hspace{.2pt}} r @{\hspace{.2pt}} r @{\hspace{.2pt}} r @{\hspace{.2pt}} r @{\hspace{.2pt}} r }
x_1  &-x_2^*  &-x_3^*& x_4  &-x_4^*&-x_3   & x_2  & x_1^*  &-x_5^* & x_6 & 0 & 0 &0 & 0     & x_6 & -x_5^* &
-x_6^* &-x_5 & 0 & 0 &0 & 0  &-x_5 & -x_6^*   & x_2  & x_1^*  & x_4  &-x_3^*&-x_3  &-x_4^* & x_1  &-x_2^*\\
x_2  & x_1^*  & x_4  &-x_3^*&-x_3  &-x_4^* & x_1  &-x_2^*  &x_6 &-x_5^* & 0 & 0 &0 & 0 &-x_5^* & x_6 &
-x_5  &-x_6^* & 0 & 0 & 0&0 & -x_6^* &-x_5   & x_1  &-x_2^*  &-x_3^*& x_4  &-x_4^*&-x_3   & x_2  & x_1^*\\
x_3  & x_4    & x_1^*& x_2^*&-x_2  & x_1   &-x_4^*& x_3^*  &0 &0 &-x_5^* & x_6 & x_6  & -x_5^* & 0 &0 &
0 & 0 & -x_6^* &-x_5 &-x_5 & -x_6^* &0  & 0  & x_4  & x_3    &-x_2  & x_1  & x_1^*& x_2^* & x_3^*&-x_4^*\\
x_4  & x_3    &-x_2  & x_1  & x_1^*& x_2^* & x_3^*&-x_4^* &0 &0 & x_6 &-x_5^* & -x_5^* & x_6 &0 &0 &
0 & 0  &-x_5& -x_6^* & -x_6^* &-x_5 & 0 & 0 & x_3  & x_4    & x_1^*& x_2^*&-x_2  & x_1   &-x_4^*& x_3^*\\
-x_4  & x_3    &-x_2  & x_1  &-x_1^*&-x_2^* &-x_3^*&-x_4^* &0 &0 & x_6  &-x_5^* & x_5^* & -x_6 &0 &0  &
0 & 0  &-x_5& -x_6^* &x_6^* & x_5 & 0 & 0   & x_3  &-x_4    & x_1^*& x_2^*& x_2  &-x_1   &-x_4^*&-x_3^*  \\
x_3  &-x_4    & x_1^*& x_2^*& x_2  &-x_1   &-x_4^*&-x_3^*  &0 &0 &-x_5^* & x_6 &-x_6  & x_5^* & 0 &0 &
0 & 0 & -x_6^* &-x_5 & x_5&x_6^* &0  & 0    & -x_4  & x_3    &-x_2  & x_1  &-x_1^*&-x_2^* &-x_3^*&-x_4^*\\
x_2  & x_1^*  &-x_4  &-x_3^*& x_3  &-x_4^* &-x_1  & x_2^* & x_6 &-x_5^* & 0 & 0 &0 & 0 &x_5^* &-x_6 &
-x_5  &-x_6^* & 0 & 0 & 0&0 &x_6^* & x_5     &x_1  &-x_2^*  &-x_3^*&-x_4  &-x_4^*&x_3    &-x_2  &-x_1^* \\
x_1  &-x_2^*  &-x_3^*&-x_4  &-x_4^*&x_3    &-x_2  &-x_1^* &-x_5^* & x_6 & 0 & 0 &0 & 0 & x_6 & x_5^*  &
-x_6^* &-x_5 & 0 & 0 &0 & 0  & x_5 &x_6^*    &  x_2  & x_1^*  &-x_4  &-x_3^*& x_3  &-x_4^* &-x_1  & x_2^*\\
x_5 & x_6 & 0 & 0 &0 & 0  & x_6 & x_5    & x_1^*  & x_2^* & x_3^*&-x_4  & x_4^* &x_3    &-x_2  & x_1 &
-x_2   & x_1   &-x_4  & x_3^*& x_3   & x_4^* & x_1^*& x_2^*  &-x_6^* & x_5^* & 0 & 0 &0 & 0  & x_5^* & -x_6^*  \\
x_6 & x_5 & 0 & 0 &0 & 0 & x_5 & x_6 &-x_2   & x_1   &-x_4  & x_3^*& x_3   & x_4^* & x_1^*& x_2^* &
x_1^*  & x_2^* & x_3^*&-x_4  & x_4^* &x_3    &-x_2  & x_1    & x_5^*  &-x_6^* & 0 & 0 & 0&0 & -x_6^* & x_5^*\\
0 &0 &x_5 & x_6 & x_6  & x_5 & 0 &0  &-x_3   &-x_4   & x_1  &-x_2^*& x_2   & x_1^* & x_4^*&-x_3^* &
-x_4   &-x_3   & x_2  & x_1^*& x_1   &-x_2^* &-x_3^*& x_4^* &0 & 0 & -x_6^* & x_5^* & x_5^* & -x_6^* &0  & 0 \\
0 &0 & x_6  & x_5 & x_5 & x_6 &0 &0  &-x_4   &-x_3   & x_2  & x_1^*& x_1   &-x_2^* &-x_3^*& x_4^* &
-x_3   &-x_4   & x_1  &-x_2^*& x_2   & x_1^* & x_4^*&-x_3^*  &0 & 0  & x_5^* & -x_6^* & -x_6^* & x_5^* & 0 & 0\\
0 &0 & x_6  & x_5 & -x_5 &-x_6&0 &0 &x_4   &-x_3   & x_2  & x_1^*&-x_1   & x_2^* & x_3^*& x_4^* &
-x_3   & x_4   & x_1  &-x_2^*&-x_2   &-x_1^* & x_4^*& x_3^*  &0 & 0  & x_5^* & -x_6^* &x_6^* &-x_5^* & 0 & 0 \\

0 &0 & x_5 & x_6 &-x_6  & -x_5 & 0 &0&-x_3   & x_4   & x_1  &-x_2^*&-x_2   &-x_1^* & x_4^*& x_3^* &
x_4   &-x_3   & x_2  & x_1^*&-x_1   & x_2^* & x_3^*& x_4^*   &0 & 0 & -x_6^* & x_5^* &-x_5^* &x_6^* &0  & 0 \\
x_6 & x_5 & 0 & 0 &0 & 0 &-x_5 &-x_6 &-x_2   & x_1   &x_4   & x_3^*&-x_3   & x_4^* &-x_1^*&-x_2^* &
x_1^* &x_2^*  & x_3^*& x_4  & x_4^* &-x_3   & x_2  &-x_1 & x_5^* &-x_6^* & 0 & 0 & 0&0 &x_6^* &-x_5^* \\
x_5 & x_6  & 0 & 0 &0 & 0 &-x_6 &-x_5& x_1^* &x_2^*  & x_3^*& x_4  & x_4^* &-x_3   & x_2  &-x_1 &
-x_2   & x_1   &x_4   & x_3^*&-x_3   & x_4^* &-x_1^*&-x_2^*    & -x_6^* & x_5^* & 0 & 0 &0 & 0  &-x_5^* & x_6^*\\
-x_6 & x_5 & 0 & 0 &0 & 0  & x_5 &- x_6 &-x_2 & x_1 &-x_4  & x_3^*& x_3   & x_4^* & x_1^*& x_2^* &
-x_1^* &- x_2^* &- x_3^*& x_4  &- x_4^* &-x_3    & x_2  &- x_1    &-x_5^* &-x_6^*  & 0 & 0 &0 & 0  &-x_6^* &-x_5^*\\
x_5  &-x_6 & 0 & 0 & 0&0 &- x_6 & x_5   & x_1^*  & x_2^* & x_3^*&-x_4  & x_4^* &x_3    &-x_2  & x_1 &
x_2   &-x_1   & x_4  &- x_3^*&- x_3   &- x_4^* &- x_1^*&-x_2^* &-x_6^* &-x_5^* & 0 & 0 &0 & 0 &-x_5^* &-x_6^* \\
0 & 0 &- x_6 & x_5 &x_5 &-x_6 &0  & 0   &-x_4   &-x_3   & x_2  & x_1^*& x_1   &-x_2^* &-x_3^*& x_4^* &
x_3   &x_4   &- x_1  &x_2^*&- x_2   &- x_1^* &- x_4^*&x_3^* & 0 &0 &-x_5^* &-x_6^* &-x_6^*  &-x_5^* & 0 &0 \\
0 & 0  & x_5&- x_6 &- x_6 & x_5 & 0 & 0 &-x_3   &-x_4   & x_1  &-x_2^*& x_2   & x_1^* & x_4^*&-x_3^* &
x_4   &x_3   &-x_2  &-x_1^*&-x_1   &x_2^* &x_3^*&- x_4^* &0 &0 &-x_6^*  &-x_5^* & x_5^* &-x_6^* &0 &0\\
0 & 0  & x_5&- x_6 & x_6 & -x_5 & 0 & 0 &-x_3   & x_4   & x_1  &-x_2^*&-x_2   &-x_1^* & x_4^*& x_3^* &
-x_4   &x_3   &- x_2  &- x_1^*&x_1   &- x_2^* &- x_3^*&- x_4^* & 0 &0 &-x_6^*  &-x_5^* & x_5^* &-x_6^* &0 &0 \\
0 & 0 &- x_6 & x_5 &0 & x_6 &-x_5  & 0  &x_4   &-x_3   & x_2  & x_1^*&-x_1   & x_2^* & x_3^*& x_4^* &
x_3   &-x_4   &-x_1  &x_2^*&x_2   &x_1^* &-x_4^*&-x_3^* & 0 &0 &-x_5^* &-x_6^* &-x_6^*  &x_5^* & 0 &0 \\
x_5  &-x_6 & 0 & 0 & 0&0 & x_6 & -x_5   & x_1^* &x_2^*  & x_3^*& x_4  & x_4^* &-x_3   & x_2  &-x_1 &
x_2   &- x_1   &-x_4   &- x_3^*&x_3   &-x_4^* &x_1^*&x_2^* &-x_6^* &-x_5^* & 0 & 0 &0 & 0 & x_5^* &-x_6^* \\
-x_6 & x_5 & 0 & 0 &0 & 0  & -x_5 & x_6  &-x_2  & x_1  &x_4  & x_3^*&-x_3   & x_4^* &-x_1^*&-x_2^* &
-x_1^* &-x_2^*  &- x_3^*&- x_4  &-x_4^* &x_3   &- x_2  &x_1    &-x_5^* &-x_6^* & 0 & 0 &0 & 0  &-x_6^* & x_5^*\\
x_2  & x_1^*  & x_4  &-x_3^*&-x_3  &-x_4^* & x_1 &-x_2^* &- x_6 & -x_5^* & 0 & 0 &0 & 0 &-x_5^* &- x_6 &
x_5 & -x_6* & 0 & 0 &0 & 0   &-x_6* & x_5   &- x_1  &x_2^*  &x_3^*&- x_4  &x_4^*&x_3   & -x_2  &- x_1^*    \\
x_1  &-x_2^*  &-x_3^*& x_4  &-x_4^*&-x_3   & x_2  & x_1^* & -x_5^* &-x_6 & 0 & 0 & 0&0 &- x_6 &-x_5^* &
-x_6* & x_5^* & 0 & 0 &0 & 0 & x_5 &-x_6*  &- x_2  &- x_1^*  &- x_4  &x_3^*&x_3  &x_4^* & -x_1  &x_2^* \\
x_4  & x_3    &-x_2  & x_1  & x_1^*& x_2^* & x_3^*&-x_4^* &0 & 0 &- x_6 &-x_5^* &-x_5^* &- x_6 &0  & 0 &
0 &0 & x_5 & -x_6* &-x_6*  & x_5 & 0 &0      &-x_3  &- x_4    &- x_1^*&- x_2^*&x_2  & -x_1   &x_4^*&- x_3^*  \\
x_3  & x_4    & x_1^*& x_2^*&-x_2  & x_1   &-x_4^*& x_3^* &0 & 0  &-x_5^*&- x_6 &- x_6 & -x_5^* & 0 & 0 &
0 &0 &-x_6*  & x_5 & x_5 & -x_6* &0 &0   &- x_4  &- x_3    &x_2  &- x_1  &- x_1^*&-x_2^* &- x_3^*&x_4^*\\
x_3  &-x_4    & x_1^*& x_2^*& x_2  &-x_1   &-x_4^*&-x_3^* &0 & 0  & -x_5^*&- x_6 & x_6 & x_5^* & 0 & 0  &
0 &0 &-x_6*  & x_5 &-x_5 & x_6* &0 &0         & x_4  &- x_3    &x_2  &- x_1  & x_1^*& x_2^* & x_3^*& x_4^* \\
-x_4  & x_3    &-x_2  & x_1  &-x_1^*&-x_2^* &-x_3^*&-x_4^*  &0 & 0 &- x_6 & -x_5^* & x_5^* & x_6 &0  & 0 &
0 &0 & x_5 &-x_6* &0  &-x_5 & x_6* &0      &-x_3  & x_4    &-x_1^*&-x_2^*&-x_2  &x_1   & x_4^*& x_3^*  \\
x_1  &-x_2^*  &-x_3^*&-x_4  &-x_4^*&x_3    &-x_2  &-x_1^* & -x_5^*  &-x_6 & 0 & 0 & 0&0 & x_6 & x_5^* &
-x_6* & x_5 & 0 & 0 &0 & 0 &-x_5 & x_6*       &- x_2  &- x_1^*  & x_4  & x_3^*&- x_3  & x_4^* & x_1  &-x_2^* \\
x_2  & x_1^*  &-x_4  &-x_3^*& x_3  &-x_4^* &-x_1  & x_2^* &- x_6 &-x_5^* & 0 & 0 &0 & 0  & x_5^* & x_6 &
x_5 & -x_6* & 0 & 0 &0 & 0  & x_6* &-x_5    &-x_1  & x_2^*  & x_3^*& x_4  & x_4^*&-x_3    & x_2  & x_1^*
\end{array}
\hspace{-2pt}
\right]
\]
}
\caption{The $[32,32,6]$ code $\mathbf{H}_5$ with fraction of zeros $\frac{1}{4}$}
\label{c1fig3}

\end{figure*}

\section{Simulation Results}
\label{sec5}
The symbol error performance of the SCODs constructed in this paper (denoted as RZCOD in the plots which means COD with Reduced number of Zeros) for $8,16,32$ and $64$ antennas are compared with that of well-known COD (denoted as SCOD) of same order in Fig. \ref{fig1} and Fig. \ref{fig2} under peak power constraint.
Similarly, Fig. \ref{figavg1} and Fig. \ref{figavg2} compare the corresponding codes under average power constraint. The average power constraint performance of RZCOD matches with that of the comparable SCOD, while the RZCOD performs  better than the corresponding SCOD under peak power constraint as seen in the figures. We also observe that the performance of our code for 8 transmit antennas matches with that of the code  $\mathbf{G}_Y$ (denoted as Yuen(8)) constructed by Yuen et al. 

\section{Discussion}
\label{sec6}
We have given construction for rate $\frac{a+1}{2^a}$ SCODs for $2^a$ antennas, for all values of $a$, with lesser number of zero entries than the known constructions. When $a+1$ is a power of 2, our construction gives SCODs with no zero entries. This case alone generalizes the constructions in \cite{YGT,TWMS,STWWWXZ,ZSXWWWT} which are only for $8$ antennas.
Some of the possible directions for further research are listed below:
\begin{itemize}
\item For arbitrary values of $a$ the fraction of zero entries in our codes is $1-\frac{a+1}{2^a} 2^{\lfloor{log_2(\frac{2^a}{a+1})}\rfloor}$. We conjecture that SCODs with smaller fraction of zero entries do not exist. It will be an interesting direction to pursue to settle this conjecture.
\item Several designs including CODs have been found useful in systems exploiting cooperative diversity. It will be interesting to investigate the suitability of the codes of this paper for cooperative diversity.
\item We have exploited the combinatorial structure of the rows of the design in \eqref{itcod} to obtain the codes with low PAPR. The interrelationship between AODs and our codes is an important direction to pursue.
\end{itemize}

\ifCLASSOPTIONcaptionsoff
  \newpage
\fi

\begin{IEEEbiography}[{\includegraphics[width=1in,height=1.25in,clip,keepaspectratio]{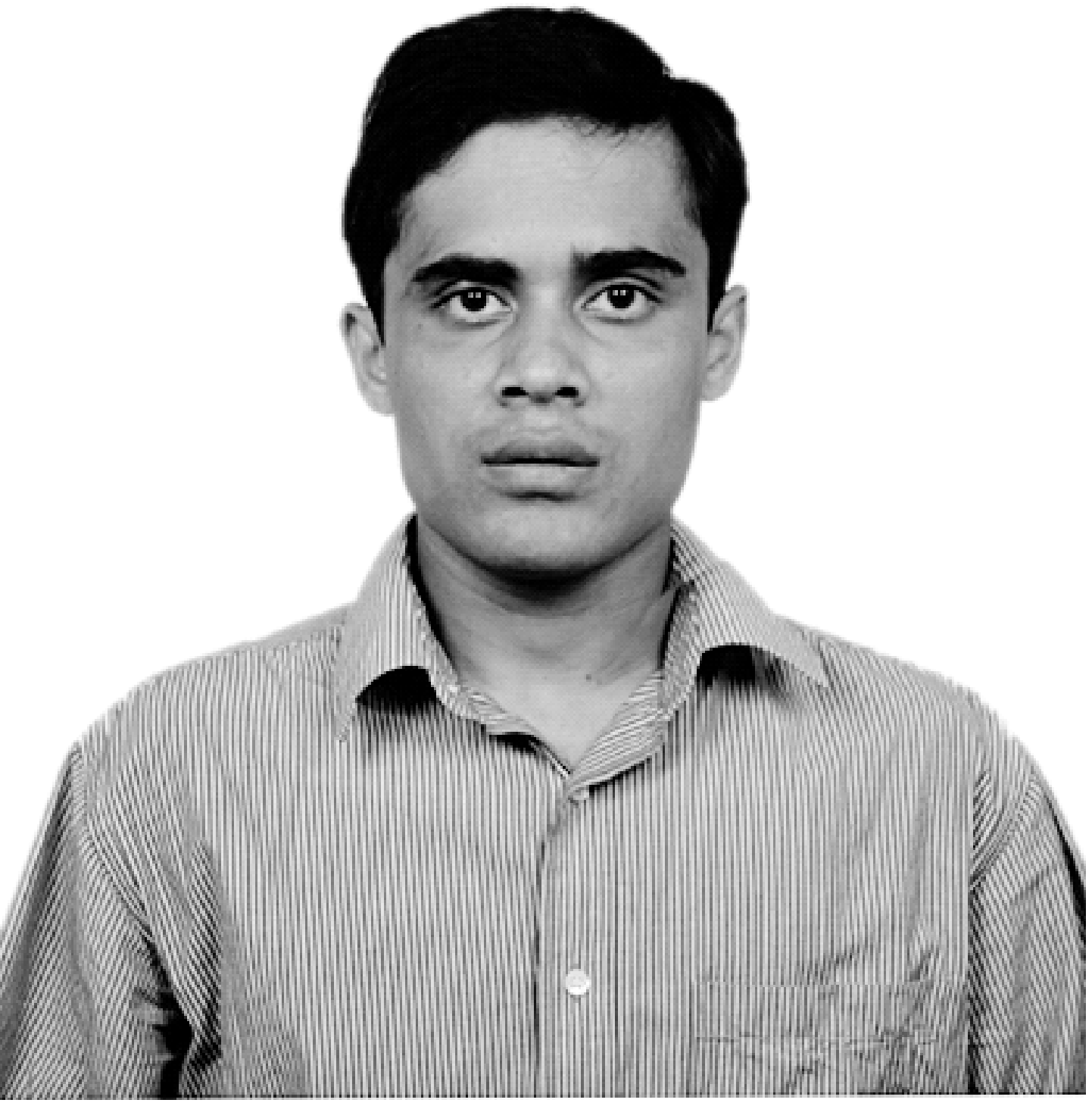}}]{Smarajit Das}
(S'2007) was born in West Bengal, India. He completed his B.E. degree at the Sardar Vallabbhai National Institute of Technology, Surat, India in 2001. He is currently a Ph.D. student in the Department of Electrical Communication Engineering, Indian Institute of Science, Bangalore, India. His primary research interests include space-time coding for MIMO channels with an emphasis on algebraic code construction techniques. 
\end{IEEEbiography}

\vfill

\begin{IEEEbiography}[{\includegraphics[width=1in,height=1.25in,clip,keepaspectratio]{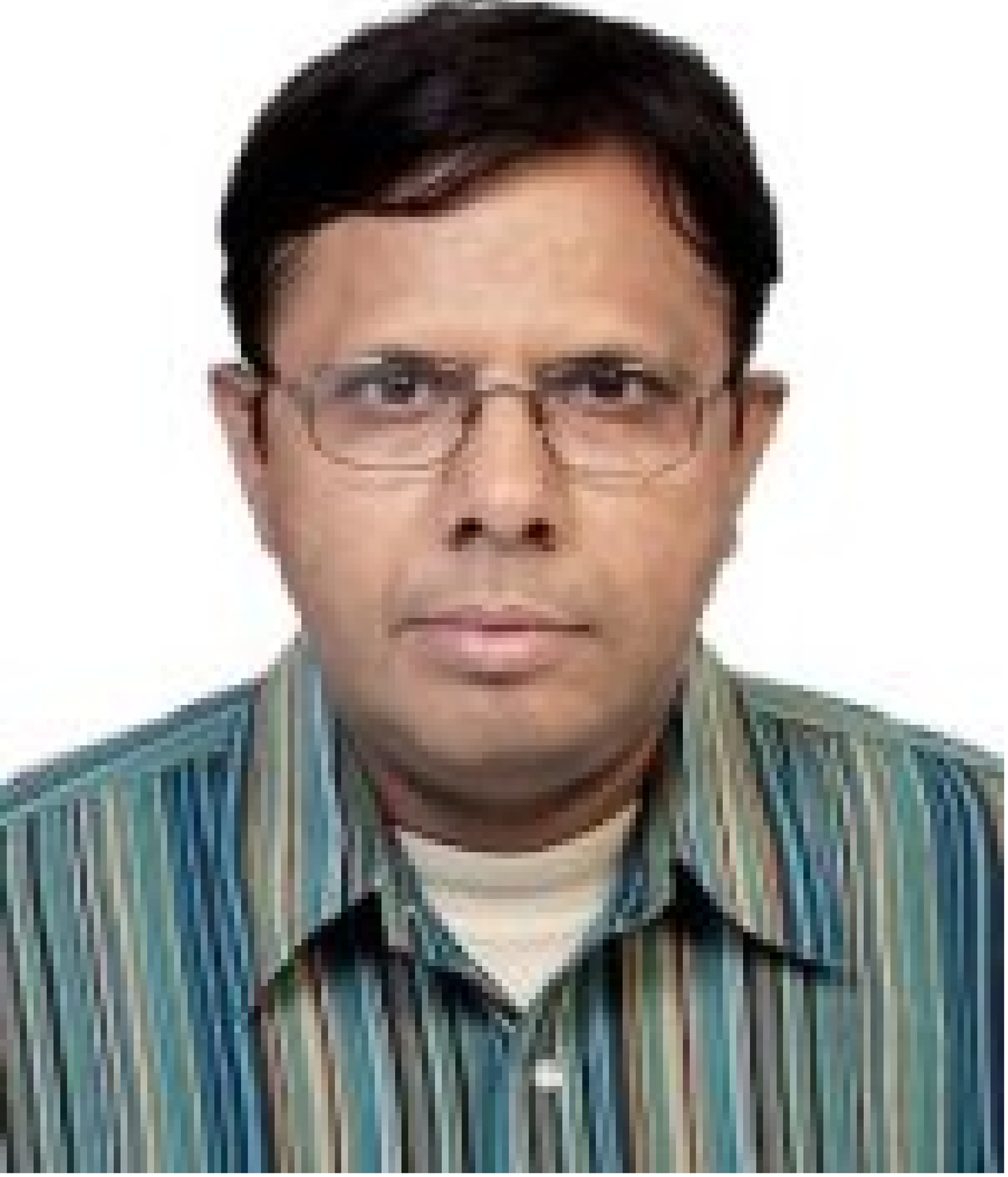}}]{B. Sundar Rajan}
(S'84-M'91-SM'98) was born in Tamil Nadu, India. He received the B.Sc. degree in mathematics from Madras University, Madras, India, the B.Tech degree in electronics from Madras Institute of Technology, Madras, and the M.Tech and Ph.D. degrees in electrical engineering from the Indian Institute of Technology, Kanpur, India, in 1979, 1982, 1984, and 1989 respectively. He was a faculty member with the Department of Electrical Engineering at the Indian Institute of Technology in Delhi, India, from 1990 to 1997. Since 1998, he has been a Professor in the Department of Electrical Communication Engineering at the Indian Institute of Science, Bangalore, India. His primary research interests are in algebraic coding, coded modulation and space-time coding.

Dr. Rajan is an Editor of IEEE Transactions on Wireless Communications from 2007 and also a Editorial Board Member of International Journal of Information and Coding Theory. He is a Fellow of Indian National Academy of Engineering and  is a Member of the American Mathematical Society. 
\end{IEEEbiography}
\vfill
\end{document}